\documentclass[aps,prl,twocolumn,showpacs,superscriptaddress,10pt]{revtex4-1}

\usepackage{amsmath}
\usepackage{amsfonts}
\usepackage{amssymb}
\usepackage{graphicx}
\usepackage{color}
\usepackage{xcolor}
\usepackage{braket}

\newcommand{\ev}[1]{\langle#1\rangle}

\DeclareRobustCommand{\orderof}{\ensuremath{\mathcal{O}}}

\newcommand{\es}[2]{\langle\hat{\sigma}_#1^#2\rangle}
\newcommand{\est}[2]{\langle\hat{\sigma}_#1^#2 (t)\rangle}
\newcommand{\ess}[2]{\langle\hat{\sigma}_#1 \hat{\sigma}_#2\rangle}
\newcommand{\esst}[4]{\langle\hat{\sigma}_#1^#2(t) \hat{\sigma}_#3^#4 (t)\rangle}

\begin{document}
\title{Single-Particle Decoherence Can Improve Spin-Squeezing Generated In Collective Dynamics}
%

\author{K. Tucker}
\affiliation{JILA, NIST, Department of Physics, University of Colorado,  Boulder, CO 80309, USA}
\affiliation{Department of Applied Mathematics, University of Colorado, Boulder, CO 80309, USA}
\author{D. Barberena}
\affiliation{JILA, NIST, Department of Physics, University of Colorado,  Boulder, CO 80309, USA}
\affiliation{Center for Theory of Quantum Matter, University of Colorado, Boulder, CO 80309, USA}
\author{R. J. Lewis-Swan}
\affiliation{JILA, NIST, Department of Physics, University of Colorado,  Boulder, CO 80309, USA}
\affiliation{Center for Theory of Quantum Matter, University of Colorado, Boulder, CO 80309, USA}

\author{J. K. Thompson}
\affiliation{JILA, NIST, Department of Physics, University of Colorado,  Boulder, CO 80309, USA}

\author{J. G. Restrepo}
\affiliation{Department of Applied Mathematics, University of Colorado, Boulder, CO 80309, USA}

\author{A. M. Rey}
\affiliation{JILA, NIST, Department of Physics, University of Colorado,  Boulder, CO 80309, USA}
\affiliation{Center for Theory of Quantum Matter, University of Colorado, Boulder, CO 80309, USA}
\begin{abstract}

We study the generation of spin-squeezing in arrays of long-lived dipoles subject to collective emission, coherent drive, elastic interactions, and spontaneous emission. Counter-intuitively, it is found that the introduction of spontaneous emission leads to an enhancement of the achievable spin-squeezing, relative to that which emerges in the steady-state of the purely collective dynamics for the same model parameters. This behavior is connected to the dynamical self-tuning of the system through a dissipative phase transition that is present in the collective system alone. Our findings will be applicable to next-generation quantum sensors harnessing correlated quantum matter, including cavity-QED and trapped ion systems.

\end{abstract}

\date{\today}
\maketitle

\noindent
\emph{Introduction --} The preparation of entangled and non-classical quantum states is a vital task for many quantum technologies, including metrology \cite{Pezze_2018} and quantum information \cite{Zeilinger_2000,Braunstein_2012}. Conventional protocols generate entanglement via coherent dynamics and seek to minimize the decoherence induced by couplings to the environment \cite{TSS_2018,Hu_2017}. However, it has been established that dissipation can itself be a powerful resource for entanglement generation under appropriate conditions. In particular, quantum reservoir engineering has established the potential to generate pure entangled steady-states by carefully tailored couplings between the system and environment \cite{Zoller_2008, Zoller_2008_2, Lukin_2013,Sorensen_2011,Polzik_2011,Wineland_2013,Zoller_1996}. 

While these engineered dissipative systems can lead to rich physics, their realization is difficult. Ultracold atoms coupled to optical cavities and trapped ion arrays are emerging as a convenient platform where both coherent and dissipative dynamics can be engineered with great controllability \cite{Norcia_2018,DPT,Leroux_2010,Davis_2019,Lev_2018,Hosten_2016,Barrett_2014,Kohler_2017,Baumann_2010,Brennecke_2007,Bohnet_2016,Arghavan_2018,Zhang_2017,Jurcevic_2017,Barreiro_2011}. In fact, these systems have garnered tremendous theoretical attention for many years \cite{Walls_1978,carm1980, drumm, drummHassan, drummCarm,diego,Morrison_2008,Li_2009,Florentin_2014,Lee_2013,Milburn_2002,Porras_2013,Lee_2014,Wolfe_2014} given the emergent new behaviors, critical phenomena, and quantum phases of matter that they can feature. For example, non-equilibrium phase transitions in collective models, featuring entangled steady states around critical points, have been identified as an appealing resource for quantum metrology \cite{Milburn_2002,Porras_2013,Lee_2014,Wolfe_2014,diego}. However, a drawback is that the timescales required to reach the steady-state are typically extremely long \cite{diego, carm1980}, specifically with respect to common experimental sources of technical noise and single-particle decoherence which are often neglected in the theoretical models. In view of this, the widely-held expectation is that single-particle decoherence will strongly limit any entanglement generated by the collective dynamics.

\begin{figure}[t!]
\includegraphics[width=8.5cm]{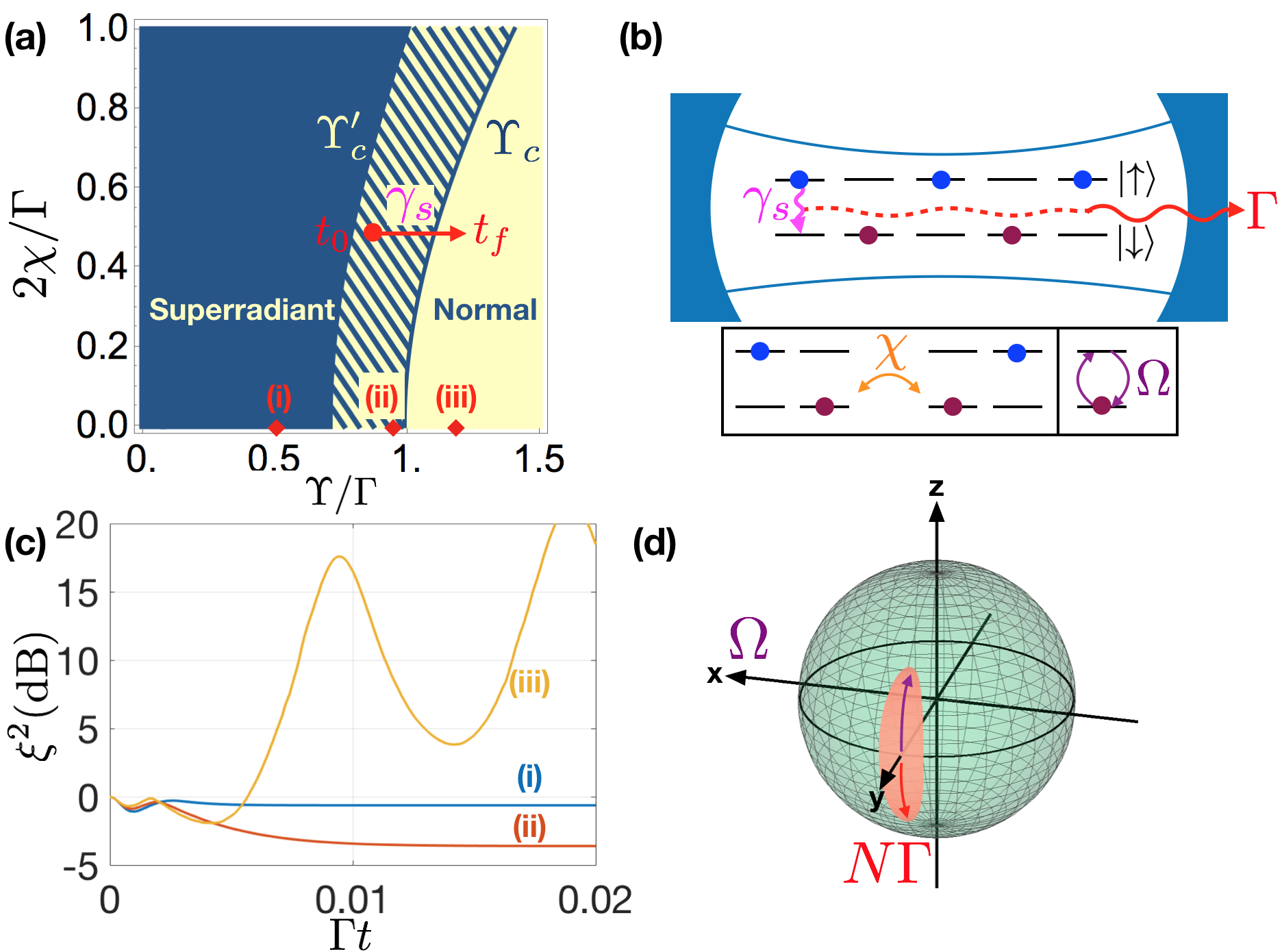}
\caption{(a) Steady-state phase diagram of an ensemble of $N$ spin-$1/2$ particles subjected to a coherent drive with Rabi frequency $\Omega = N\Upsilon/2$, collective emission at rate $\Gamma$,  collective spin-exchange interactions $\chi$, and single-particle spontaneous emission at rate $\gamma_s$. This system can be engineered using an optical cavity or trapped ion arrays (b). In the absence of spontaneous emission, the system undergoes a non-equilibrium phase transition (superradiant to normal)  signaled by a change in the total steady-state atomic inversion, which serves as an order parameter. Approaching the transition point from the superradiant phase [points (i) and (ii)], the coherent drive (in the $\hat{x}$-direction) and collective emission combine to generate spin-squeezing along $\hat{x}$, as shown in (c) and (d). In the normal phase no squeezing is observed [point (iii)]. For all three graphs in panel (c), $N = 2000$ and all spins are initially polarized along $-\hat{x}$. Panel (d) explicitly displays the Bloch sphere overlaid with a squeezed collective spin distribution of the steady-state (pink). Note that this is for illustrative purposes, and that the actual position and orientation of the squeezing can vary with parameters. Introducing finite $\gamma_s$ allows the system to dynamically traverse the phase-diagram [red arrow in (a)] and enhances the achievable spin-squeezing in the striped region of panel (a). 
}
\label{fig1}
\end{figure}

Here, we demonstrate that the introduction of single-particle decoherence in a collective open system is not necessarily detrimental and in fact can lead to improved entanglement relative to the steady-state of the collective dynamics with the same model parameters. Specifically, we show that in a system subject to collective drive and dissipation, the spin-squeezing achievable in the purely collective steady-state at fixed parameters can be improved by advantageous transient dynamics induced by single-particle spontaneous emission or dephasing.

The mechanism driving this phenomenon is the destruction of collective coherence due to single-particle decoherence, which dynamically reduces the effective particle number, allowing the system to dynamically traverse the corresponding non-equilibrium phase-diagram [see Fig.~\ref{fig1}(a)], and in turn access regimes that display large transient squeezing. While our analysis of this phenomenon is framed from a cavity-QED perspective, we note that similar conclusions can be drawn in more general models including arrays of trapped ions \cite{Shankar_2017,Milburn_2002} and superconducting qubits \cite{Fink,Mlynek2014}.

\noindent
\emph{Model --} We consider an ensemble of $N$ atoms in an  optical lattice supported by a standing wave optical cavity, illustrated in Fig.~\ref{fig1}(b). A single common mode of the cavity couples two internal states of the atoms, $\ket{\uparrow}$ and $\ket{\downarrow}$, which encode a spin-$1/2$ degree of freedom. To realize coherent driving of the dipoles the cavity is pumped with an external coherent field that is resonant with the atomic transition, and upon adiabatic elimination of the intracavity field \cite{Haake_1971} (which we assume evolves rapidly compared to relevant timescales) the dynamics of the atomic degrees of freedom can be described by a master equation for the atomic density operator $\hat{\rho}$ \cite{diego}
\begin{flalign}
\label{eqn_master}
\frac{\partial\hat{\rho}}{\partial t} &= -\frac{i}{\hbar}[\hat{H},\hat{\rho}] + L_c[\hat{\rho}] + L_s[\hat{\rho}], \\
\hat{H} &= \hbar\chi \hat{J}_+\hat{J}_- + \hbar\Omega \hat{J}_x,
\end{flalign}
where $\hat{J}_\alpha = \sum_{i=1}^N\frac{1}{2}\hat{\sigma}_i^\alpha$ for $\alpha = x,y,z$, $\hat{\sigma}_i^\alpha$ are the Pauli operators on the Hilbert space for each spin $i = 1, 2, ...,N$, and $\hat{J}_\pm = \hat{J}_x \pm i\hat{J}_y$ are collective raising and lowering operators. The first term in $\hat{H}$ corresponds to a collective exchange interaction realized by detuning the cavity from the atomic transition and characterized by $\chi$, and the second to a coherent drive characterized by $\Omega$. The dissipative part of Eq.~(\ref{eqn_master}) includes a collective decay term with rate $\Gamma$ given by $L_c[\hat{\rho}] = \Gamma L(\hat{J}_-)[\hat{\rho}]$ and single-particle spontaneous emission at rate $\gamma_s$ given by $L_s[\hat{\rho}] = \gamma_s \sum_{i=1}^N L(\hat{\sigma}_i^-)[\hat{\rho}]$, where the Lindblad superoperator is $L(\hat{O})[\hat{\rho}] = \hat{O}\hat{\rho}\hat{O}^\dagger - \{\hat{O}^\dagger \hat{O}, \hat{\rho}\}/2$ for a given operator $\hat{O}$. The former can arise due to leakage of the intracavity field via the mirrors, while the latter is a result of the finite lifetime of the excited state of the transition. Other types of single-particle decoherence (e.g., dephasing) would result in similar behavior, so we only consider spontaneous emission here.

\noindent
\emph{Collective physics --} Before discussing the effects of spontaneous emission, we review the behavior of the collective system when $\gamma_s = 0$. As the dynamics is entirely described by collective operators, then the total spin operator $\hat{J}^2 = \hat{J}_x^2 + \hat{J}_y^2 + \hat{J}_z^2$ is conserved during evolution. Consequently, if we restrict ourselves to initializing the atoms in a coherent spin state \cite{Radcliffe_1971}, which is an eigenstate of $\hat{J}^2$ with eigenvalue $J(J+1)$ with $J=N/2$, then the available Hilbert space in which the dynamics and steady-state exist is greatly reduced to only $N+1$ states (relative to $2^N$ for $N$ spin-$1/2$s). With this simplification, an analytic solution is available for the steady-state density operator $\hat{\rho}_{ss}$ \cite{diego, carm1980} from which all relevant collective spin observables can be computed. Previous work \cite{carm1980,diego} has demonstrated that as a function of  $\Upsilon\equiv \Big(2 \Omega /N\Big)$ and for large $N$, the steady state exhibits a non-equilibrium second-order phase transition in the thermodynamic limit, described by an abrupt change in behavior of  the order parameter $\ev{\hat{J}_z}$ at a critical value given by:
\begin{equation}
    \label{eqn_omega_c}
   \Upsilon_c = \sqrt{\Gamma^2 + 4\chi^2}.
\end{equation} 
The critical point separates a superradiant phase for $\Upsilon< \Upsilon_c$ characterized by non-zero inversion $\vert\ev{\hat{J}_z} \vert>0 $, and a normal phase for $\Upsilon >\Upsilon_c$ with zero inversion $\ev{\hat{J}_z} = 0$ [see Fig.~\ref{fig1}(a)]. The critical point  $\Upsilon_c$ also delineates regions in the phase diagram for which the steady-state of the atomic ensemble is spin-squeezed. The squeezing is characterized by the parameter \cite{Wineland_1992}
\begin{equation}
    \xi^2 = \min_{{\bf n}_\perp}\frac{N (\Delta \hat{J}_{{\bf n}_\perp})^2}{|\ev{{\bf \hat{J}}}|^2},
\end{equation}
where $\ev{\hat{{\bf J}}} = (\ev{\hat{J}_x}, \ev{\hat{J}_y}, \ev{\hat{J}_z})$ defines the collective Bloch vector, ${\bf n}_\perp$ is a unit vector orthogonal to $\ev{\hat{{\bf J}}}$, and ($\Delta \hat{J}_{{\bf n}_\perp})^2 = \ev{(\hat{{\bf J}} \cdot {\bf n}_\perp)^2} - \ev{\hat{{\bf J}} \cdot {\bf n}_\perp}^2$ is the variance of the collective spin operator in the direction of ${\bf n}_\perp$. Squeezing, $\xi^2 < 1$, is an entanglement witness and quantifies the utility of the spin state for quantum sensing applications \cite{KU}. 

Figure \ref{fig1}(c) illustrates that just below  $\Upsilon_c$ the steady-state is squeezed, due to the finely balanced competition of the coherent drive and the nonlinear dynamics induced by the collective dissipation [see Fig.~\ref{fig1}(d)]. Specifically, as $\Upsilon$ approaches the threshold $\Upsilon_c$ from below [curves (i) and (ii)] the system relaxes into an increasingly squeezed state with $\xi^2 < 1$. However, for $\Upsilon > \Upsilon_c$ the squeezing is abruptly lost beyond an early transient. It should be noted that for $\Upsilon < \Upsilon_c$ a careful selection of initial conditions becomes necessary to reach the squeezed steady-state quickly, and to avoid an oscillatory phase known to exist near the critical point when $\vert\chi\vert > 0$ \cite{diego}.

\begin{figure}[t!]
    \includegraphics[width=8.5cm]{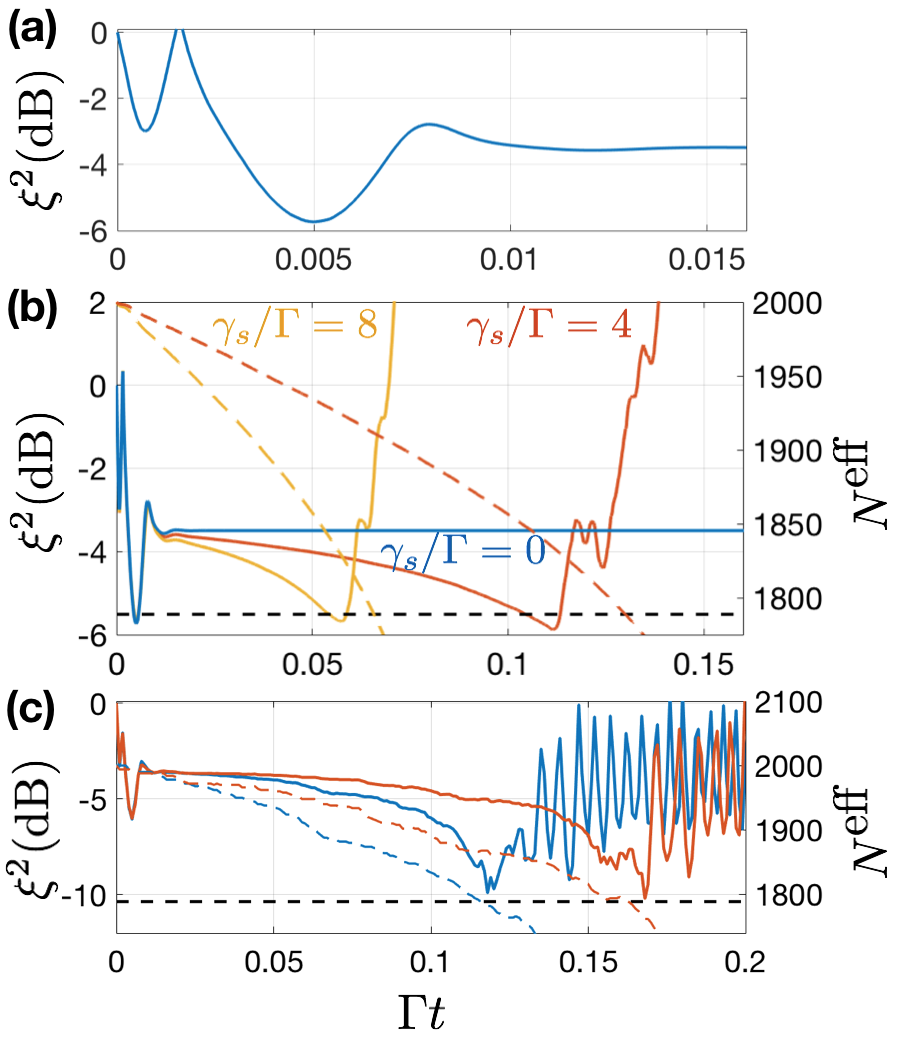}
    \caption{(a) Squeezing versus time for $N = 2000$, $\chi/\Gamma = 1$, $\Upsilon/\Upsilon_c = 0.9$, and $\gamma_s/\Gamma = 0$ showing squeezing early in the dynamics. (b) Examples of enhanced transient squeezing for $\gamma_s/\Gamma \neq 0$ compared to collective case ($\gamma_s/\Gamma = 0$). Solid lines indicate squeezing $\xi^2(t)$ and dashed lines the corresponding time-dependent effective system size $N^{\mathrm{eff}}(t)$ (matching colors). (c) Squeezing $\xi^2(t)$ (solid) and effective system size $N^{\mathrm{eff}}(t)$ (dashed) computed from two individual trajectories of the numerical method with $\gamma_ s/\Gamma = 4$. The horizontal black line corresponds to the critical effective particle number $N_c = 2\Omega/\Upsilon_c$ for which the transition between superradiant and normal phases occurs. In each panel, all spins are initially polarized along $-\hat{x}$.}
    \label{fig3}
\end{figure}

\noindent
\emph{Effects of single-particle dissipation --} When $\gamma_s \neq 0$, the collective $\hat{J}^2$ symmetry is broken by the single-particle decoherence. This means the dynamics are free to explore a larger portion of the full Hilbert space of $2^N$ states, compared to the limited $N+1$ states of the collective model. Due to this increased complexity, an analytic formula for the steady-state is not available. However, a mean-field analysis can give useful insight into the steady-state phase diagram of the system, including the position of critical transitions and transient behavior. These predictions can be confirmed by efficient numerical simulation \cite{mcwf, mcwf2, mcwf3} of the full quantum dynamics described by the master equation [Eq.~(\ref{eqn_master})], which also allows us to investigate quantum features such as spin squeezing. 
In the mean-field approximation, we generate equations of motion for the expectation values $\est{i}{\alpha}$ (identical for all particles due to permutational symmetry) from Eq.~(\ref{eqn_master}) and assume that all higher-order expectations factorize, i.e. $\esst{i}{\alpha}{j}{\beta} = \est{i}{\alpha}\est{j}{\beta}$ for $i\neq j$ \cite{SM}.

The mean-field analysis indicates that many of the qualitative features of the collective physics, particularly the steady-state behavior, remain when single-particle spontaneous emission is included. 
Specifically, for $\gamma_s \neq 0$ there is a critical point $\Upsilon_c' \equiv \Upsilon_c/\sqrt{2}$ delineating superradiant and normal phases characterized by the long-time limit of collective observables. Moreover, numerical simulations of the full quantum dynamics reveal that $\Upsilon_c'$ also marks the boundary between a squeezed steady state in the superradiant phase and the absence of long-time squeezing in the normal phase \cite{SM}. This transition is illustrated in Fig.~\ref{fig1}(a).


\noindent
\emph{Enhanced squeezing --} We now turn our focus to a quantitative analysis of the effects of spontaneous emission on the achievable spin-squeezing, both in the steady-state and in the transient dynamics. Naively, one might expect that single-particle dissipation only leads to a degradation of the squeezing generated by the collective dynamics \cite{Shammah_2018}. 
Between $\Upsilon_c' < \Upsilon < \Upsilon_c$ [striped region (ii) Fig.~\ref{fig1}(a)] we find appreciable squeezing develops in the transient dynamics on a time-scale for which both collective and single-particle effects are relevant. In particular, the predicted squeezing exceeds what is seen in the collective steady-state for $\gamma_s = 0$. 

Figure \ref{fig3}(b) illustrates the spin-squeezing dynamics in this region $\Upsilon_c' < \Upsilon< \Upsilon_c$ for several values of $\gamma_s/\Gamma$. We observe two distinct timescales where squeezing occurs. The first is short-lived and occurs early in the dynamics [see also Fig.~\ref{fig3}(a)], where collective behavior dominates and the squeezing is independent of $\gamma_s/\Gamma$. A second transient occurs at timescales $\sim 1/\gamma_s$ and is intrinsically connected to single-particle effects. We find the achievable squeezing is enhanced with respect to the purely collective steady-state ($\gamma_s = 0$) and so focus on understanding the cause of this transient behavior in the following.


\noindent
\emph{Squeezing mechanism --} The transient  squeezing for $\Upsilon_c' < \Upsilon< \Upsilon_c$ occurring  at timescales $\sim 1/\gamma_s$ can be understood within the framework of the collective steady-state. Specifically, the enhancement can be understood as a subtle consequence of the destruction of collective coherence by single-particle dissipation. We argue that reducing the collective coherence leads to an effective increase of $\Upsilon$ over time  [$\Upsilon\to\Upsilon^{\mbox{eff}}(t)$], which  allows the system to dynamically traverse the collective phase-diagram into regions with higher spin-squeezing. 


Our argument is illustrated by plotting in Fig.~\ref{fig3}(b) the time-evolution of an effective atom-number related to the total spin $\hat{J}^2$ as $N^{\mbox{eff}}(t) \equiv 2\sqrt{(1/4) + \ev{\hat{J^2}}(t)} - 1$. For $\gamma_s = 0$ we have $N^{\mbox{eff}}(t) = N$, but for $\gamma_s > 0$ we observe the effective system size decays, $N^{\mbox{eff}}(t) \leq N$, making  $\Upsilon^{\mbox{eff}}(t)\equiv 2\Omega/N^{\mbox{eff}}(t)$ grow over time even though the coherent drive remains constant.

One can therefore  dynamically approach and even cross the critical point  $\Upsilon_c$  as  the system evolves. This is confirmed by the strong correlation between the timing of the crossing of the threshold atom number, $N^{\mbox{eff}}(t^*)$ [determined from $\Upsilon^{\mbox{eff}}(t^*)=\Upsilon_c$ and indicated by a dashed horizontal line in Fig.~\ref{fig3}(b)]  and the loss of squeezing  for $t>t^*$ for the  different $\gamma_s$. This provides evidence that squeezing is dynamically lost as the system effectively transitions from the superradiant to normal phases, corresponding to the crossover from squeezed to unsqueezed regimes in the collective model.

However, in Fig.~\ref{fig3}(b), we observe a small quantitative disagreement between these timescales. To confirm the idea that squeezing disappears as a result of $\Upsilon^{\mbox{eff}}$ crossing to the normal phase and  demonstrate that the observed deviation in the averaged quantities is a result of quantum noise, in Fig.~\ref{fig3}(c) we perform investigation of individual trajectories,  which mimic a typical experimental realization. We use these trajectories  to simulate the open system dynamics via a Monte-Carlo wavefunction method which  unravels the evolution of the density matrix [Eq.~(\ref{eqn_master})] into an ensemble of pure state wavefunctions evolving accordingly to a non-Hermitian Hamiltonian. Dissipation is further incorporated within each of these independent trajectories by stochastic jumps that project the wavefunction \cite{SM}. As shown in Figure \ref{fig3}(c), in a single trajectory squeezing features an abrupt change in behavior exactly for   $t>t^*$. However, as visible in Fig.~\ref{fig3}, the fact that  $t^*$ varies from trajectory to trajectory  explains, on the one hand, the   moderate  discrepancy in timescales mentioned  above  and,  on the other, the  net reduction on the  optimal observed   squeezing  when  an  average over many trajectories is taken. The later is necessary to recover the master equation results. Related to this last point, we note that these results indicate that sources of technical noise, such as shot-to-shot fluctuations in atom number will need to be kept sufficiently small (\textit{i.e.}, sub-Poissonian) as they can also lead to a smearing out of the crossover into the superradiant phase and reduce the achievable squeezing overall.


\begin{figure}[t!]
\includegraphics[width=8cm]{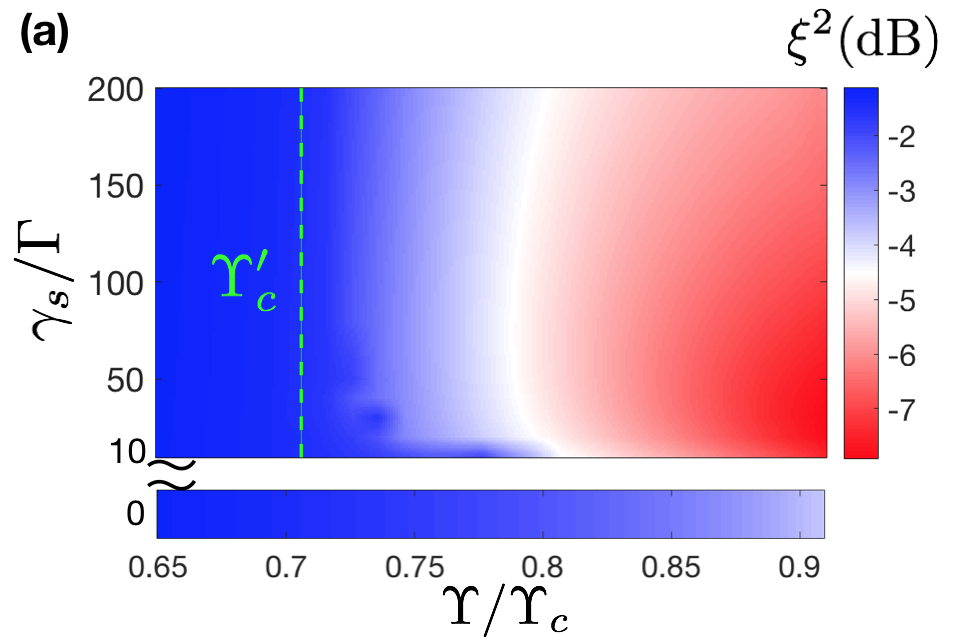}
\includegraphics[width=8cm]{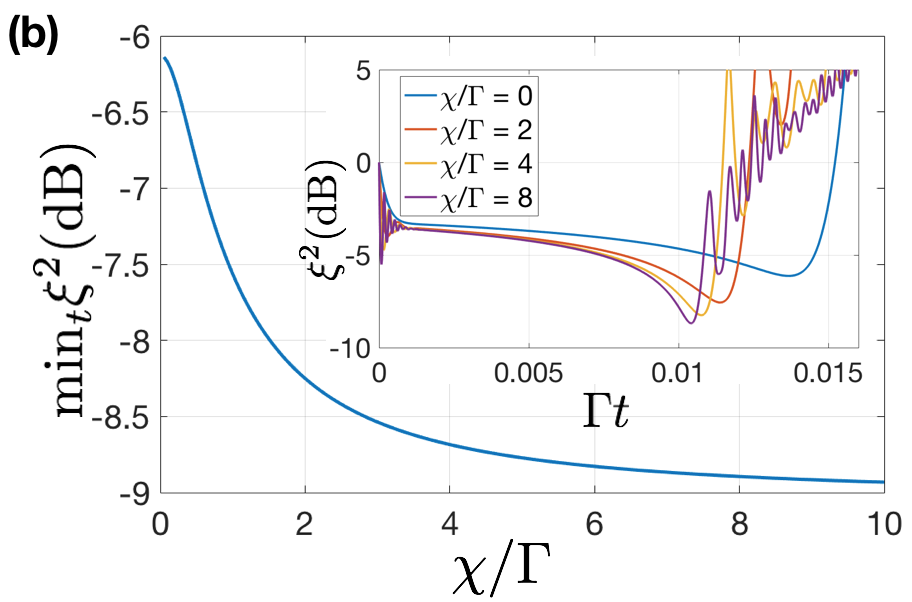}
\caption{(a) Minimum transient spin-squeezing (see text for clarification) as a function of normalized drive amplitude $\Upsilon/\Upsilon_c$ and spontaneous emission rate $\gamma_s/\Gamma$ with $\chi/\Gamma = 1$. The strip below the main panel shows a magnified view of the $\gamma_s/\Gamma = 0$ result for comparison. Note the break in the vertical axis, which is required since attainable simulation times cannot capture the squeezing behavior occurring on timescales of $1/\gamma_s$ when this value is very large. (b) Minimum transient spin-squeezing as a function of the interaction strength $\chi/\Gamma$ with fixed $\Upsilon/\Upsilon_c = 0.9$ and $\gamma_s/\Gamma = 50$. Inset: Squeezing versus time for a selection of values of $\chi/\Gamma$ and same $\Upsilon,\gamma_s$ as main panel. For (a) and (b) we compute the dynamics using a truncated cumulant expansion \cite{SM} and $N = 10^4$. Initial conditions in (a) are the coherent spin state (CSS) in the $-\hat{x}$-direction, and in (b) are taken to be the CSS in the direction of the mean-field steady-state (for each $\chi/\Gamma$ and $\Upsilon$) when $\gamma_s/\Gamma = 0$ to account for the rotations that result from different values of $\chi/\Gamma$.}
\label{fig4}
\end{figure}

In Fig.~\ref{fig4}(a) we investigate the minimum squeezing obtained in the transient dynamics, after the initial collective minimum, as a function of the normalized drive amplitude $\Upsilon$ and spontaneous emission rate $\gamma_s/\Gamma$. The introduction of spontaneous emission ($\gamma_s/\Gamma \neq 0$) clearly improves the attainable squeezing within the region of $\Upsilon_c^{\prime} < \Upsilon < \Upsilon_c$ relative to the collective case ($\gamma_s/\Gamma = 0$, shown in the lower strip). This improvement occurs for even relatively small values of $\gamma_s$, although as $\gamma_s$ is increased the best transient squeezing, attained for $\Upsilon$ approaching $\Upsilon_c$, gradually degrades. On the other hand, for $\Upsilon < \Upsilon_c^{\prime}$ a stable steady-state is quickly reached and squeezing is not enhanced by introducing single-particle decoherence.

While our qualitative understanding of the mechanism driving the enhanced squeezing has so far not required discussion of the collective exchange interactions, the best achievable squeezing does quantitatively depend on $\chi$ for $\Upsilon < \Upsilon_c^{\prime}$. This is demonstrated in Fig.~\ref{fig4}(b), where we plot the minimum transient squeezing as a function of $\chi/\Gamma$ for $\gamma_s/\Gamma = 50$. It is apparent that increasing the interaction strength $\chi$ leads to an appreciable improvement in the optimal squeezing, particularly in the region $0 < \chi/\Gamma \lesssim 2$. However, the inset of Fig.~\ref{fig4}(b) indicates that an increased interaction strength does not significantly change the qualitative dynamics of the squeezing, beyond obtaining the optimal value at earlier times.

\noindent
\emph{Experimental realization and outlook --} 
The spin model we have discussed could be realized by coupling an optical cavity to the narrow linewidth optical clock transitions available in alkaline earth atoms \cite{Norcia_2018,DPT}. We require that $\kappa \gg g\sqrt{N}$ and $\kappa \gg \gamma_s$ (bad cavity limit) with $2g$ the single photon Rabi frequency and $\kappa$ the cavity linewidth, to ensure that the intracavity field can be adiabatically eliminated and thus realize the desired  spin model [Eqs.~(1) and (2)]. In this limit, spin-spin interactions can be engineered by detuning the cavity from the atomic transition by $\Delta_c$ which leads to a tunable interaction strength $\chi = 4g^2\Delta_c/(4\Delta_c^2 + \kappa^2)$ \cite{Norcia_2018}. Similarly, the collective dissipation arises due to photon leakage and is characterized by $\Gamma = 4g^2\kappa/(4\Delta_c^2 + \kappa^2)$ \cite{Norcia_2018,Norcia_2016}. These features have previously been demonstrated using both the $^1$S$_0$-$^3$P$_0$ transition in $^{87}$Sr \cite{Norcia_2018,Norcia_2016} and $^1$S$_0$-$^3$P$_1$ transition in $^{88}$Sr \cite{DPT}. The former has a natural linewidth of $\gamma \approx 2\pi \times 1$~ mHz and $2g = 2\pi\times 8$~Hz \cite{Norcia_2018}. State-of-the-art AMO experiments have demonstrated coherence of the $^1$S$_0$-$^3$P$_0$ transition of up to $1/\gamma_s \approx 10$~s \cite{Hutson2019} which corresponds to $\gamma_s/\Gamma_s \approx 200$ for $\Delta_c = \kappa = 2\pi\times 150$~kHz and thus $\chi/\Gamma \approx 1$. For $N \sim 10^4$ atoms, dissipatively enhanced squeezing of $\xi^2 \approx 9$~dB is then in principle achievable on timescales $t\sim 2$~s. 
A similar implementation can also be realized in trapped ion arrays, where a pseudospin-$1/2$ is encoded in the hyperfine states of the ion. In this case, resonant microwaves can be used to coherently directly drive the spins \cite{Arghavan_2018}. The collective interactions proportional to $\chi$ can be  engineered via coupling to a common motional mode of the ion crystal \cite{Bohnet_2016}, and collective dissipation can be implemented by coupling to a second ionic species \cite{Shankar_2017}.

In summary, we have identified an intriguing and experimentally relevant situation where relatively large single-particle decoherence can enhance spin-squeezing as long as collective decoherence remains the dominant dissipative process. We expect our results to have immediate applications for quantum metrology, specifically in the generation of squeezing on long-lived optical transitions for next-generation optical atomic clocks, whilst also being relevant for quantum simulation.

\begin{acknowledgments}
We are grateful for feedback from M.~A. Norcia  and J. T. Young on the manuscript. This work is supported by the AFOSR grant FA9550-18-1-0319 and its MURI Initiative, by the DARPA and ARO grant W911NF-16-1-0576, the ARO single investigator award W911NF-19-1-0210,  the NSF PHY1820885, NSF JILA-PFC PHY-1734006 grants, and by NIST. 
\end{acknowledgments}

\bibliography{enhanced}

\begin{thebibliography}{55}%
\makeatletter
\providecommand \@ifxundefined [1]{%
 \@ifx{#1\undefined}
}%
\providecommand \@ifnum [1]{%
 \ifnum #1\expandafter \@firstoftwo
 \else \expandafter \@secondoftwo
 \fi
}%
\providecommand \@ifx [1]{%
 \ifx #1\expandafter \@firstoftwo
 \else \expandafter \@secondoftwo
 \fi
}%
\providecommand \natexlab [1]{#1}%
\providecommand \enquote  [1]{``#1''}%
\providecommand \bibnamefont  [1]{#1}%
\providecommand \bibfnamefont [1]{#1}%
\providecommand \citenamefont [1]{#1}%
\providecommand \href@noop [0]{\@secondoftwo}%
\providecommand \href [0]{\begingroup \@sanitize@url \@href}%
\providecommand \@href[1]{\@@startlink{#1}\@@href}%
\providecommand \@@href[1]{\endgroup#1\@@endlink}%
\providecommand \@sanitize@url [0]{\catcode `\\12\catcode `\$12\catcode
  `\&12\catcode `\#12\catcode `\^12\catcode `\_12\catcode `\%12\relax}%
\providecommand \@@startlink[1]{}%
\providecommand \@@endlink[0]{}%
\providecommand \url  [0]{\begingroup\@sanitize@url \@url }%
\providecommand \@url [1]{\endgroup\@href {#1}{\urlprefix }}%
\providecommand \urlprefix  [0]{URL }%
\providecommand \Eprint [0]{\href }%
\providecommand \doibase [0]{http://dx.doi.org/}%
\providecommand \selectlanguage [0]{\@gobble}%
\providecommand \bibinfo  [0]{\@secondoftwo}%
\providecommand \bibfield  [0]{\@secondoftwo}%
\providecommand \translation [1]{[#1]}%
\providecommand \BibitemOpen [0]{}%
\providecommand \bibitemStop [0]{}%
\providecommand \bibitemNoStop [0]{.\EOS\space}%
\providecommand \EOS [0]{\spacefactor3000\relax}%
\providecommand \BibitemShut  [1]{\csname bibitem#1\endcsname}%
\let\auto@bib@innerbib\@empty
\bibitem [{\citenamefont {Pezz\`e}\ \emph {et~al.}(2018)\citenamefont
  {Pezz\`e}, \citenamefont {Smerzi}, \citenamefont {Oberthaler}, \citenamefont
  {Schmied},\ and\ \citenamefont {Treutlein}}]{Pezze_2018}%
  \BibitemOpen
  \bibfield  {author} {\bibinfo {author} {\bibfnamefont {L.}~\bibnamefont
  {Pezz\`e}}, \bibinfo {author} {\bibfnamefont {A.}~\bibnamefont {Smerzi}},
  \bibinfo {author} {\bibfnamefont {M.~K.}\ \bibnamefont {Oberthaler}},
  \bibinfo {author} {\bibfnamefont {R.}~\bibnamefont {Schmied}}, \ and\
  \bibinfo {author} {\bibfnamefont {P.}~\bibnamefont {Treutlein}},\ }\href
  {\doibase 10.1103/RevModPhys.90.035005} {\bibfield  {journal} {\bibinfo
  {journal} {Rev. Mod. Phys.}\ }\textbf {\bibinfo {volume} {90}},\ \bibinfo
  {pages} {035005} (\bibinfo {year} {2018})}\BibitemShut {NoStop}%
\bibitem [{\citenamefont {Bouwmeester}\ \emph {et~al.}(2000)\citenamefont
  {Bouwmeester}, \citenamefont {Ekert},\ and\ \citenamefont
  {Zeilinger}}]{Zeilinger_2000}%
  \BibitemOpen
  \bibinfo {editor} {\bibfnamefont {D.}~\bibnamefont {Bouwmeester}}, \bibinfo
  {editor} {\bibfnamefont {A.}~\bibnamefont {Ekert}}, \ and\ \bibinfo {editor}
  {\bibfnamefont {A.}~\bibnamefont {Zeilinger}},\ eds.,\ \href {\doibase
  10.1007/978-3-662-04209-0} {\emph {\bibinfo {title} {The Physics of Quantum
  Information}}}\ (\bibinfo  {publisher} {Springer Berlin Heidelberg},\
  \bibinfo {year} {2000})\BibitemShut {NoStop}%
\bibitem [{\citenamefont {Braunstein}\ and\ \citenamefont
  {Pati}(2012)}]{Braunstein_2012}%
  \BibitemOpen
  \bibfield  {author} {\bibinfo {author} {\bibfnamefont {S.}~\bibnamefont
  {Braunstein}}\ and\ \bibinfo {author} {\bibfnamefont {A.}~\bibnamefont
  {Pati}},\ }\href {https://books.google.com/books?id=PTfpCAAAQBAJ} {\emph
  {\bibinfo {title} {Quantum Information with Continuous Variables}}}\
  (\bibinfo  {publisher} {Springer Netherlands},\ \bibinfo {year}
  {2012})\BibitemShut {NoStop}%
\bibitem [{\citenamefont {Lewis-Swan}\ \emph {et~al.}(2018)\citenamefont
  {Lewis-Swan}, \citenamefont {Norcia}, \citenamefont {Cline}, \citenamefont
  {Thompson},\ and\ \citenamefont {Rey}}]{TSS_2018}%
  \BibitemOpen
  \bibfield  {author} {\bibinfo {author} {\bibfnamefont {R.~J.}\ \bibnamefont
  {Lewis-Swan}}, \bibinfo {author} {\bibfnamefont {M.~A.}\ \bibnamefont
  {Norcia}}, \bibinfo {author} {\bibfnamefont {J.~R.~K.}\ \bibnamefont
  {Cline}}, \bibinfo {author} {\bibfnamefont {J.~K.}\ \bibnamefont {Thompson}},
  \ and\ \bibinfo {author} {\bibfnamefont {A.~M.}\ \bibnamefont {Rey}},\ }\href
  {\doibase 10.1103/PhysRevLett.121.070403} {\bibfield  {journal} {\bibinfo
  {journal} {Phys. Rev. Lett.}\ }\textbf {\bibinfo {volume} {121}},\ \bibinfo
  {pages} {070403} (\bibinfo {year} {2018})}\BibitemShut {NoStop}%
\bibitem [{\citenamefont {Hu}\ \emph {et~al.}(2017)\citenamefont {Hu},
  \citenamefont {Chen}, \citenamefont {Vendeiro}, \citenamefont {Urvoy},
  \citenamefont {Braverman},\ and\ \citenamefont {Vuleti\ifmmode~\acute{c}\else
  \'{c}\fi{}}}]{Hu_2017}%
  \BibitemOpen
  \bibfield  {author} {\bibinfo {author} {\bibfnamefont {J.}~\bibnamefont
  {Hu}}, \bibinfo {author} {\bibfnamefont {W.}~\bibnamefont {Chen}}, \bibinfo
  {author} {\bibfnamefont {Z.}~\bibnamefont {Vendeiro}}, \bibinfo {author}
  {\bibfnamefont {A.}~\bibnamefont {Urvoy}}, \bibinfo {author} {\bibfnamefont
  {B.}~\bibnamefont {Braverman}}, \ and\ \bibinfo {author} {\bibfnamefont
  {V.}~\bibnamefont {Vuleti\ifmmode~\acute{c}\else \'{c}\fi{}}},\ }\href
  {\doibase 10.1103/PhysRevA.96.050301} {\bibfield  {journal} {\bibinfo
  {journal} {Phys. Rev. A}\ }\textbf {\bibinfo {volume} {96}},\ \bibinfo
  {pages} {050301} (\bibinfo {year} {2017})}\BibitemShut {NoStop}%
\bibitem [{\citenamefont {Kraus}\ \emph {et~al.}(2008)\citenamefont {Kraus},
  \citenamefont {B\"uchler}, \citenamefont {Diehl}, \citenamefont {Kantian},
  \citenamefont {Micheli},\ and\ \citenamefont {Zoller}}]{Zoller_2008}%
  \BibitemOpen
  \bibfield  {author} {\bibinfo {author} {\bibfnamefont {B.}~\bibnamefont
  {Kraus}}, \bibinfo {author} {\bibfnamefont {H.~P.}\ \bibnamefont
  {B\"uchler}}, \bibinfo {author} {\bibfnamefont {S.}~\bibnamefont {Diehl}},
  \bibinfo {author} {\bibfnamefont {A.}~\bibnamefont {Kantian}}, \bibinfo
  {author} {\bibfnamefont {A.}~\bibnamefont {Micheli}}, \ and\ \bibinfo
  {author} {\bibfnamefont {P.}~\bibnamefont {Zoller}},\ }\href {\doibase
  10.1103/PhysRevA.78.042307} {\bibfield  {journal} {\bibinfo  {journal} {Phys.
  Rev. A}\ }\textbf {\bibinfo {volume} {78}},\ \bibinfo {pages} {042307}
  (\bibinfo {year} {2008})}\BibitemShut {NoStop}%
\bibitem [{\citenamefont {Diehl}\ \emph {et~al.}(2008)\citenamefont {Diehl},
  \citenamefont {Micheli}, \citenamefont {Kantian}, \citenamefont {Kraus},
  \citenamefont {B{\"u}chler},\ and\ \citenamefont {Zoller}}]{Zoller_2008_2}%
  \BibitemOpen
  \bibfield  {author} {\bibinfo {author} {\bibfnamefont {S.}~\bibnamefont
  {Diehl}}, \bibinfo {author} {\bibfnamefont {A.}~\bibnamefont {Micheli}},
  \bibinfo {author} {\bibfnamefont {A.}~\bibnamefont {Kantian}}, \bibinfo
  {author} {\bibfnamefont {B.}~\bibnamefont {Kraus}}, \bibinfo {author}
  {\bibfnamefont {H.~P.}\ \bibnamefont {B{\"u}chler}}, \ and\ \bibinfo {author}
  {\bibfnamefont {P.}~\bibnamefont {Zoller}},\ }\href
  {https://doi.org/10.1038/nphys1073} {\bibfield  {journal} {\bibinfo
  {journal} {Nature Physics}\ }\textbf {\bibinfo {volume} {4}},\ \bibinfo
  {pages} {878 EP } (\bibinfo {year} {2008})}\BibitemShut {NoStop}%
\bibitem [{\citenamefont {Dalla~Torre}\ \emph {et~al.}(2013)\citenamefont
  {Dalla~Torre}, \citenamefont {Otterbach}, \citenamefont {Demler},
  \citenamefont {Vuletic},\ and\ \citenamefont {Lukin}}]{Lukin_2013}%
  \BibitemOpen
  \bibfield  {author} {\bibinfo {author} {\bibfnamefont {E.~G.}\ \bibnamefont
  {Dalla~Torre}}, \bibinfo {author} {\bibfnamefont {J.}~\bibnamefont
  {Otterbach}}, \bibinfo {author} {\bibfnamefont {E.}~\bibnamefont {Demler}},
  \bibinfo {author} {\bibfnamefont {V.}~\bibnamefont {Vuletic}}, \ and\
  \bibinfo {author} {\bibfnamefont {M.~D.}\ \bibnamefont {Lukin}},\ }\href
  {\doibase 10.1103/PhysRevLett.110.120402} {\bibfield  {journal} {\bibinfo
  {journal} {Phys. Rev. Lett.}\ }\textbf {\bibinfo {volume} {110}},\ \bibinfo
  {pages} {120402} (\bibinfo {year} {2013})}\BibitemShut {NoStop}%
\bibitem [{\citenamefont {Kastoryano}\ \emph {et~al.}(2011)\citenamefont
  {Kastoryano}, \citenamefont {Reiter},\ and\ \citenamefont
  {S\o{}rensen}}]{Sorensen_2011}%
  \BibitemOpen
  \bibfield  {author} {\bibinfo {author} {\bibfnamefont {M.~J.}\ \bibnamefont
  {Kastoryano}}, \bibinfo {author} {\bibfnamefont {F.}~\bibnamefont {Reiter}},
  \ and\ \bibinfo {author} {\bibfnamefont {A.~S.}\ \bibnamefont
  {S\o{}rensen}},\ }\href {\doibase 10.1103/PhysRevLett.106.090502} {\bibfield
  {journal} {\bibinfo  {journal} {Phys. Rev. Lett.}\ }\textbf {\bibinfo
  {volume} {106}},\ \bibinfo {pages} {090502} (\bibinfo {year}
  {2011})}\BibitemShut {NoStop}%
\bibitem [{\citenamefont {Krauter}\ \emph {et~al.}(2011)\citenamefont
  {Krauter}, \citenamefont {Muschik}, \citenamefont {Jensen}, \citenamefont
  {Wasilewski}, \citenamefont {Petersen}, \citenamefont {Cirac},\ and\
  \citenamefont {Polzik}}]{Polzik_2011}%
  \BibitemOpen
  \bibfield  {author} {\bibinfo {author} {\bibfnamefont {H.}~\bibnamefont
  {Krauter}}, \bibinfo {author} {\bibfnamefont {C.~A.}\ \bibnamefont
  {Muschik}}, \bibinfo {author} {\bibfnamefont {K.}~\bibnamefont {Jensen}},
  \bibinfo {author} {\bibfnamefont {W.}~\bibnamefont {Wasilewski}}, \bibinfo
  {author} {\bibfnamefont {J.~M.}\ \bibnamefont {Petersen}}, \bibinfo {author}
  {\bibfnamefont {J.~I.}\ \bibnamefont {Cirac}}, \ and\ \bibinfo {author}
  {\bibfnamefont {E.~S.}\ \bibnamefont {Polzik}},\ }\href {\doibase
  10.1103/PhysRevLett.107.080503} {\bibfield  {journal} {\bibinfo  {journal}
  {Phys. Rev. Lett.}\ }\textbf {\bibinfo {volume} {107}},\ \bibinfo {pages}
  {080503} (\bibinfo {year} {2011})}\BibitemShut {NoStop}%
\bibitem [{\citenamefont {Lin}\ \emph {et~al.}(2013)\citenamefont {Lin},
  \citenamefont {Gaebler}, \citenamefont {Reiter}, \citenamefont {Tan},
  \citenamefont {Bowler}, \citenamefont {S{\o}rensen}, \citenamefont
  {Leibfried},\ and\ \citenamefont {Wineland}}]{Wineland_2013}%
  \BibitemOpen
  \bibfield  {author} {\bibinfo {author} {\bibfnamefont {Y.}~\bibnamefont
  {Lin}}, \bibinfo {author} {\bibfnamefont {J.~P.}\ \bibnamefont {Gaebler}},
  \bibinfo {author} {\bibfnamefont {F.}~\bibnamefont {Reiter}}, \bibinfo
  {author} {\bibfnamefont {T.~R.}\ \bibnamefont {Tan}}, \bibinfo {author}
  {\bibfnamefont {R.}~\bibnamefont {Bowler}}, \bibinfo {author} {\bibfnamefont
  {A.~S.}\ \bibnamefont {S{\o}rensen}}, \bibinfo {author} {\bibfnamefont
  {D.}~\bibnamefont {Leibfried}}, \ and\ \bibinfo {author} {\bibfnamefont
  {D.~J.}\ \bibnamefont {Wineland}},\ }\href {\doibase 10.1038/nature12801}
  {\bibfield  {journal} {\bibinfo  {journal} {Nature}\ }\textbf {\bibinfo
  {volume} {504}},\ \bibinfo {pages} {415} (\bibinfo {year}
  {2013})}\BibitemShut {NoStop}%
\bibitem [{\citenamefont {Poyatos}\ \emph {et~al.}(1996)\citenamefont
  {Poyatos}, \citenamefont {Cirac},\ and\ \citenamefont
  {Zoller}}]{Zoller_1996}%
  \BibitemOpen
  \bibfield  {author} {\bibinfo {author} {\bibfnamefont {J.~F.}\ \bibnamefont
  {Poyatos}}, \bibinfo {author} {\bibfnamefont {J.~I.}\ \bibnamefont {Cirac}},
  \ and\ \bibinfo {author} {\bibfnamefont {P.}~\bibnamefont {Zoller}},\ }\href
  {\doibase 10.1103/PhysRevLett.77.4728} {\bibfield  {journal} {\bibinfo
  {journal} {Phys. Rev. Lett.}\ }\textbf {\bibinfo {volume} {77}},\ \bibinfo
  {pages} {4728} (\bibinfo {year} {1996})}\BibitemShut {NoStop}%
\bibitem [{\citenamefont {Norcia}\ \emph {et~al.}(2018)\citenamefont {Norcia},
  \citenamefont {Lewis-Swan}, \citenamefont {Cline}, \citenamefont {Zhu},
  \citenamefont {Rey},\ and\ \citenamefont {Thompson}}]{Norcia_2018}%
  \BibitemOpen
  \bibfield  {author} {\bibinfo {author} {\bibfnamefont {M.~A.}\ \bibnamefont
  {Norcia}}, \bibinfo {author} {\bibfnamefont {R.~J.}\ \bibnamefont
  {Lewis-Swan}}, \bibinfo {author} {\bibfnamefont {J.~R.~K.}\ \bibnamefont
  {Cline}}, \bibinfo {author} {\bibfnamefont {B.}~\bibnamefont {Zhu}}, \bibinfo
  {author} {\bibfnamefont {A.~M.}\ \bibnamefont {Rey}}, \ and\ \bibinfo
  {author} {\bibfnamefont {J.~K.}\ \bibnamefont {Thompson}},\ }\href {\doibase
  10.1126/science.aar3102} {\bibfield  {journal} {\bibinfo  {journal}
  {Science}\ }\textbf {\bibinfo {volume} {361}},\ \bibinfo {pages} {259}
  (\bibinfo {year} {2018})}\BibitemShut {NoStop}%
\bibitem [{\citenamefont {Muniz}\ \emph {et~al.}(2019)\citenamefont {Muniz},
  \citenamefont {Barberena}, \citenamefont {Lewis-Swan}, \citenamefont {Young},
  \citenamefont {Cline}, \citenamefont {Rey},\ and\ \citenamefont
  {Thompson}}]{DPT}%
  \BibitemOpen
  \bibfield  {author} {\bibinfo {author} {\bibfnamefont {J.~A.}\ \bibnamefont
  {Muniz}}, \bibinfo {author} {\bibfnamefont {D.}~\bibnamefont {Barberena}},
  \bibinfo {author} {\bibfnamefont {R.~J.}\ \bibnamefont {Lewis-Swan}},
  \bibinfo {author} {\bibfnamefont {D.~J.}\ \bibnamefont {Young}}, \bibinfo
  {author} {\bibfnamefont {J.~R.~K.}\ \bibnamefont {Cline}}, \bibinfo {author}
  {\bibfnamefont {A.~M.}\ \bibnamefont {Rey}}, \ and\ \bibinfo {author}
  {\bibfnamefont {J.~K.}\ \bibnamefont {Thompson}},\ }\href@noop {} {\enquote
  {\bibinfo {title} {Exploring non-equilibrium phases of matter with a million
  long-lived optical dipoles in a cavity},}\ } (\bibinfo {year} {2019}),\
  \Eprint {http://arxiv.org/abs/arXiv:1910.00439} {arXiv:1910.00439}
  \BibitemShut {NoStop}%
\bibitem [{\citenamefont {Leroux}\ \emph {et~al.}(2010)\citenamefont {Leroux},
  \citenamefont {Schleier-Smith},\ and\ \citenamefont
  {Vuleti\ifmmode~\acute{c}\else \'{c}\fi{}}}]{Leroux_2010}%
  \BibitemOpen
  \bibfield  {author} {\bibinfo {author} {\bibfnamefont {I.~D.}\ \bibnamefont
  {Leroux}}, \bibinfo {author} {\bibfnamefont {M.~H.}\ \bibnamefont
  {Schleier-Smith}}, \ and\ \bibinfo {author} {\bibfnamefont {V.}~\bibnamefont
  {Vuleti\ifmmode~\acute{c}\else \'{c}\fi{}}},\ }\href {\doibase
  10.1103/PhysRevLett.104.073602} {\bibfield  {journal} {\bibinfo  {journal}
  {Phys. Rev. Lett.}\ }\textbf {\bibinfo {volume} {104}},\ \bibinfo {pages}
  {073602} (\bibinfo {year} {2010})}\BibitemShut {NoStop}%
\bibitem [{\citenamefont {Davis}\ \emph {et~al.}(2019)\citenamefont {Davis},
  \citenamefont {Bentsen}, \citenamefont {Homeier}, \citenamefont {Li},\ and\
  \citenamefont {Schleier-Smith}}]{Davis_2019}%
  \BibitemOpen
  \bibfield  {author} {\bibinfo {author} {\bibfnamefont {E.~J.}\ \bibnamefont
  {Davis}}, \bibinfo {author} {\bibfnamefont {G.}~\bibnamefont {Bentsen}},
  \bibinfo {author} {\bibfnamefont {L.}~\bibnamefont {Homeier}}, \bibinfo
  {author} {\bibfnamefont {T.}~\bibnamefont {Li}}, \ and\ \bibinfo {author}
  {\bibfnamefont {M.~H.}\ \bibnamefont {Schleier-Smith}},\ }\href {\doibase
  10.1103/PhysRevLett.122.010405} {\bibfield  {journal} {\bibinfo  {journal}
  {Phys. Rev. Lett.}\ }\textbf {\bibinfo {volume} {122}},\ \bibinfo {pages}
  {010405} (\bibinfo {year} {2019})}\BibitemShut {NoStop}%
\bibitem [{\citenamefont {Vaidya}\ \emph {et~al.}(2018)\citenamefont {Vaidya},
  \citenamefont {Guo}, \citenamefont {Kroeze}, \citenamefont {Ballantine},
  \citenamefont {Koll\'ar}, \citenamefont {Keeling},\ and\ \citenamefont
  {Lev}}]{Lev_2018}%
  \BibitemOpen
  \bibfield  {author} {\bibinfo {author} {\bibfnamefont {V.~D.}\ \bibnamefont
  {Vaidya}}, \bibinfo {author} {\bibfnamefont {Y.}~\bibnamefont {Guo}},
  \bibinfo {author} {\bibfnamefont {R.~M.}\ \bibnamefont {Kroeze}}, \bibinfo
  {author} {\bibfnamefont {K.~E.}\ \bibnamefont {Ballantine}}, \bibinfo
  {author} {\bibfnamefont {A.~J.}\ \bibnamefont {Koll\'ar}}, \bibinfo {author}
  {\bibfnamefont {J.}~\bibnamefont {Keeling}}, \ and\ \bibinfo {author}
  {\bibfnamefont {B.~L.}\ \bibnamefont {Lev}},\ }\href {\doibase
  10.1103/PhysRevX.8.011002} {\bibfield  {journal} {\bibinfo  {journal} {Phys.
  Rev. X}\ }\textbf {\bibinfo {volume} {8}},\ \bibinfo {pages} {011002}
  (\bibinfo {year} {2018})}\BibitemShut {NoStop}%
\bibitem [{\citenamefont {Hosten}\ \emph {et~al.}(2016)\citenamefont {Hosten},
  \citenamefont {Krishnakumar}, \citenamefont {Engelsen},\ and\ \citenamefont
  {Kasevich}}]{Hosten_2016}%
  \BibitemOpen
  \bibfield  {author} {\bibinfo {author} {\bibfnamefont {O.}~\bibnamefont
  {Hosten}}, \bibinfo {author} {\bibfnamefont {R.}~\bibnamefont
  {Krishnakumar}}, \bibinfo {author} {\bibfnamefont {N.~J.}\ \bibnamefont
  {Engelsen}}, \ and\ \bibinfo {author} {\bibfnamefont {M.~A.}\ \bibnamefont
  {Kasevich}},\ }\href {\doibase 10.1126/science.aaf3397} {\bibfield  {journal}
  {\bibinfo  {journal} {Science}\ }\textbf {\bibinfo {volume} {352}},\ \bibinfo
  {pages} {1552} (\bibinfo {year} {2016})}\BibitemShut {NoStop}%
\bibitem [{\citenamefont {Baden}\ \emph {et~al.}(2014)\citenamefont {Baden},
  \citenamefont {Arnold}, \citenamefont {Grimsmo}, \citenamefont {Parkins},\
  and\ \citenamefont {Barrett}}]{Barrett_2014}%
  \BibitemOpen
  \bibfield  {author} {\bibinfo {author} {\bibfnamefont {M.~P.}\ \bibnamefont
  {Baden}}, \bibinfo {author} {\bibfnamefont {K.~J.}\ \bibnamefont {Arnold}},
  \bibinfo {author} {\bibfnamefont {A.~L.}\ \bibnamefont {Grimsmo}}, \bibinfo
  {author} {\bibfnamefont {S.}~\bibnamefont {Parkins}}, \ and\ \bibinfo
  {author} {\bibfnamefont {M.~D.}\ \bibnamefont {Barrett}},\ }\href {\doibase
  10.1103/PhysRevLett.113.020408} {\bibfield  {journal} {\bibinfo  {journal}
  {Phys. Rev. Lett.}\ }\textbf {\bibinfo {volume} {113}},\ \bibinfo {pages}
  {020408} (\bibinfo {year} {2014})}\BibitemShut {NoStop}%
\bibitem [{\citenamefont {Kohler}\ \emph {et~al.}(2017)\citenamefont {Kohler},
  \citenamefont {Spethmann}, \citenamefont {Schreppler},\ and\ \citenamefont
  {Stamper-Kurn}}]{Kohler_2017}%
  \BibitemOpen
  \bibfield  {author} {\bibinfo {author} {\bibfnamefont {J.}~\bibnamefont
  {Kohler}}, \bibinfo {author} {\bibfnamefont {N.}~\bibnamefont {Spethmann}},
  \bibinfo {author} {\bibfnamefont {S.}~\bibnamefont {Schreppler}}, \ and\
  \bibinfo {author} {\bibfnamefont {D.~M.}\ \bibnamefont {Stamper-Kurn}},\
  }\href {\doibase 10.1103/PhysRevLett.118.063604} {\bibfield  {journal}
  {\bibinfo  {journal} {Phys. Rev. Lett.}\ }\textbf {\bibinfo {volume} {118}},\
  \bibinfo {pages} {063604} (\bibinfo {year} {2017})}\BibitemShut {NoStop}%
\bibitem [{\citenamefont {Baumann}\ \emph {et~al.}(2010)\citenamefont
  {Baumann}, \citenamefont {Guerlin}, \citenamefont {Brennecke},\ and\
  \citenamefont {Esslinger}}]{Baumann_2010}%
  \BibitemOpen
  \bibfield  {author} {\bibinfo {author} {\bibfnamefont {K.}~\bibnamefont
  {Baumann}}, \bibinfo {author} {\bibfnamefont {C.}~\bibnamefont {Guerlin}},
  \bibinfo {author} {\bibfnamefont {F.}~\bibnamefont {Brennecke}}, \ and\
  \bibinfo {author} {\bibfnamefont {T.}~\bibnamefont {Esslinger}},\ }\href
  {\doibase 10.1038/nature09009} {\bibfield  {journal} {\bibinfo  {journal}
  {Nature}\ }\textbf {\bibinfo {volume} {464}},\ \bibinfo {pages} {1301}
  (\bibinfo {year} {2010})}\BibitemShut {NoStop}%
\bibitem [{\citenamefont {Brennecke}\ \emph {et~al.}(2007)\citenamefont
  {Brennecke}, \citenamefont {Donner}, \citenamefont {Ritter}, \citenamefont
  {Bourdel}, \citenamefont {K\"{o}hl},\ and\ \citenamefont
  {Esslinger}}]{Brennecke_2007}%
  \BibitemOpen
  \bibfield  {author} {\bibinfo {author} {\bibfnamefont {F.}~\bibnamefont
  {Brennecke}}, \bibinfo {author} {\bibfnamefont {T.}~\bibnamefont {Donner}},
  \bibinfo {author} {\bibfnamefont {S.}~\bibnamefont {Ritter}}, \bibinfo
  {author} {\bibfnamefont {T.}~\bibnamefont {Bourdel}}, \bibinfo {author}
  {\bibfnamefont {M.}~\bibnamefont {K\"{o}hl}}, \ and\ \bibinfo {author}
  {\bibfnamefont {T.}~\bibnamefont {Esslinger}},\ }\href {\doibase
  10.1038/nature06120} {\bibfield  {journal} {\bibinfo  {journal} {Nature}\
  }\textbf {\bibinfo {volume} {450}},\ \bibinfo {pages} {268} (\bibinfo {year}
  {2007})}\BibitemShut {NoStop}%
\bibitem [{\citenamefont {Bohnet}\ \emph {et~al.}(2016)\citenamefont {Bohnet},
  \citenamefont {Sawyer}, \citenamefont {Britton}, \citenamefont {Wall},
  \citenamefont {Rey}, \citenamefont {Foss-Feig},\ and\ \citenamefont
  {Bollinger}}]{Bohnet_2016}%
  \BibitemOpen
  \bibfield  {author} {\bibinfo {author} {\bibfnamefont {J.~G.}\ \bibnamefont
  {Bohnet}}, \bibinfo {author} {\bibfnamefont {B.~C.}\ \bibnamefont {Sawyer}},
  \bibinfo {author} {\bibfnamefont {J.~W.}\ \bibnamefont {Britton}}, \bibinfo
  {author} {\bibfnamefont {M.~L.}\ \bibnamefont {Wall}}, \bibinfo {author}
  {\bibfnamefont {A.~M.}\ \bibnamefont {Rey}}, \bibinfo {author} {\bibfnamefont
  {M.}~\bibnamefont {Foss-Feig}}, \ and\ \bibinfo {author} {\bibfnamefont
  {J.~J.}\ \bibnamefont {Bollinger}},\ }\href {\doibase
  10.1126/science.aad9958} {\bibfield  {journal} {\bibinfo  {journal}
  {Science}\ }\textbf {\bibinfo {volume} {352}},\ \bibinfo {pages} {1297}
  (\bibinfo {year} {2016})}\BibitemShut {NoStop}%
\bibitem [{\citenamefont {Safavi-Naini}\ \emph {et~al.}(2018)\citenamefont
  {Safavi-Naini}, \citenamefont {Lewis-Swan}, \citenamefont {Bohnet},
  \citenamefont {G\"arttner}, \citenamefont {Gilmore}, \citenamefont {Jordan},
  \citenamefont {Cohn}, \citenamefont {Freericks}, \citenamefont {Rey},\ and\
  \citenamefont {Bollinger}}]{Arghavan_2018}%
  \BibitemOpen
  \bibfield  {author} {\bibinfo {author} {\bibfnamefont {A.}~\bibnamefont
  {Safavi-Naini}}, \bibinfo {author} {\bibfnamefont {R.~J.}\ \bibnamefont
  {Lewis-Swan}}, \bibinfo {author} {\bibfnamefont {J.~G.}\ \bibnamefont
  {Bohnet}}, \bibinfo {author} {\bibfnamefont {M.}~\bibnamefont {G\"arttner}},
  \bibinfo {author} {\bibfnamefont {K.~A.}\ \bibnamefont {Gilmore}}, \bibinfo
  {author} {\bibfnamefont {J.~E.}\ \bibnamefont {Jordan}}, \bibinfo {author}
  {\bibfnamefont {J.}~\bibnamefont {Cohn}}, \bibinfo {author} {\bibfnamefont
  {J.~K.}\ \bibnamefont {Freericks}}, \bibinfo {author} {\bibfnamefont {A.~M.}\
  \bibnamefont {Rey}}, \ and\ \bibinfo {author} {\bibfnamefont {J.~J.}\
  \bibnamefont {Bollinger}},\ }\href {\doibase 10.1103/PhysRevLett.121.040503}
  {\bibfield  {journal} {\bibinfo  {journal} {Phys. Rev. Lett.}\ }\textbf
  {\bibinfo {volume} {121}},\ \bibinfo {pages} {040503} (\bibinfo {year}
  {2018})}\BibitemShut {NoStop}%
\bibitem [{\citenamefont {Zhang}\ \emph {et~al.}(2017)\citenamefont {Zhang},
  \citenamefont {Pagano}, \citenamefont {Hess}, \citenamefont {Kyprianidis},
  \citenamefont {Becker}, \citenamefont {Kaplan}, \citenamefont {Gorshkov},
  \citenamefont {Gong},\ and\ \citenamefont {Monroe}}]{Zhang_2017}%
  \BibitemOpen
  \bibfield  {author} {\bibinfo {author} {\bibfnamefont {J.}~\bibnamefont
  {Zhang}}, \bibinfo {author} {\bibfnamefont {G.}~\bibnamefont {Pagano}},
  \bibinfo {author} {\bibfnamefont {P.~W.}\ \bibnamefont {Hess}}, \bibinfo
  {author} {\bibfnamefont {A.}~\bibnamefont {Kyprianidis}}, \bibinfo {author}
  {\bibfnamefont {P.}~\bibnamefont {Becker}}, \bibinfo {author} {\bibfnamefont
  {H.}~\bibnamefont {Kaplan}}, \bibinfo {author} {\bibfnamefont {A.~V.}\
  \bibnamefont {Gorshkov}}, \bibinfo {author} {\bibfnamefont {Z.-X.}\
  \bibnamefont {Gong}}, \ and\ \bibinfo {author} {\bibfnamefont
  {C.}~\bibnamefont {Monroe}},\ }\href {\doibase 10.1038/nature24654}
  {\bibfield  {journal} {\bibinfo  {journal} {Nature}\ }\textbf {\bibinfo
  {volume} {551}},\ \bibinfo {pages} {601} (\bibinfo {year}
  {2017})}\BibitemShut {NoStop}%
\bibitem [{\citenamefont {Jurcevic}\ \emph {et~al.}(2017)\citenamefont
  {Jurcevic}, \citenamefont {Shen}, \citenamefont {Hauke}, \citenamefont
  {Maier}, \citenamefont {Brydges}, \citenamefont {Hempel}, \citenamefont
  {Lanyon}, \citenamefont {Heyl}, \citenamefont {Blatt},\ and\ \citenamefont
  {Roos}}]{Jurcevic_2017}%
  \BibitemOpen
  \bibfield  {author} {\bibinfo {author} {\bibfnamefont {P.}~\bibnamefont
  {Jurcevic}}, \bibinfo {author} {\bibfnamefont {H.}~\bibnamefont {Shen}},
  \bibinfo {author} {\bibfnamefont {P.}~\bibnamefont {Hauke}}, \bibinfo
  {author} {\bibfnamefont {C.}~\bibnamefont {Maier}}, \bibinfo {author}
  {\bibfnamefont {T.}~\bibnamefont {Brydges}}, \bibinfo {author} {\bibfnamefont
  {C.}~\bibnamefont {Hempel}}, \bibinfo {author} {\bibfnamefont {B.~P.}\
  \bibnamefont {Lanyon}}, \bibinfo {author} {\bibfnamefont {M.}~\bibnamefont
  {Heyl}}, \bibinfo {author} {\bibfnamefont {R.}~\bibnamefont {Blatt}}, \ and\
  \bibinfo {author} {\bibfnamefont {C.~F.}\ \bibnamefont {Roos}},\ }\href
  {\doibase 10.1103/PhysRevLett.119.080501} {\bibfield  {journal} {\bibinfo
  {journal} {Phys. Rev. Lett.}\ }\textbf {\bibinfo {volume} {119}},\ \bibinfo
  {pages} {080501} (\bibinfo {year} {2017})}\BibitemShut {NoStop}%
\bibitem [{\citenamefont {Barreiro}\ \emph {et~al.}(2011)\citenamefont
  {Barreiro}, \citenamefont {M\"{u}ller}, \citenamefont {Schindler},
  \citenamefont {Nigg}, \citenamefont {Monz}, \citenamefont {Chwalla},
  \citenamefont {Hennrich}, \citenamefont {Roos}, \citenamefont {Zoller},\ and\
  \citenamefont {Blatt}}]{Barreiro_2011}%
  \BibitemOpen
  \bibfield  {author} {\bibinfo {author} {\bibfnamefont {J.~T.}\ \bibnamefont
  {Barreiro}}, \bibinfo {author} {\bibfnamefont {M.}~\bibnamefont
  {M\"{u}ller}}, \bibinfo {author} {\bibfnamefont {P.}~\bibnamefont
  {Schindler}}, \bibinfo {author} {\bibfnamefont {D.}~\bibnamefont {Nigg}},
  \bibinfo {author} {\bibfnamefont {T.}~\bibnamefont {Monz}}, \bibinfo {author}
  {\bibfnamefont {M.}~\bibnamefont {Chwalla}}, \bibinfo {author} {\bibfnamefont
  {M.}~\bibnamefont {Hennrich}}, \bibinfo {author} {\bibfnamefont {C.~F.}\
  \bibnamefont {Roos}}, \bibinfo {author} {\bibfnamefont {P.}~\bibnamefont
  {Zoller}}, \ and\ \bibinfo {author} {\bibfnamefont {R.}~\bibnamefont
  {Blatt}},\ }\href {\doibase 10.1038/nature09801} {\bibfield  {journal}
  {\bibinfo  {journal} {Nature}\ }\textbf {\bibinfo {volume} {470}},\ \bibinfo
  {pages} {486} (\bibinfo {year} {2011})}\BibitemShut {NoStop}%
\bibitem [{\citenamefont {Walls}\ \emph {et~al.}(1978)\citenamefont {Walls},
  \citenamefont {Drummond}, \citenamefont {Hassan},\ and\ \citenamefont
  {Carmichael}}]{Walls_1978}%
  \BibitemOpen
  \bibfield  {author} {\bibinfo {author} {\bibfnamefont {D.~F.}\ \bibnamefont
  {Walls}}, \bibinfo {author} {\bibfnamefont {P.~D.}\ \bibnamefont {Drummond}},
  \bibinfo {author} {\bibfnamefont {S.~S.}\ \bibnamefont {Hassan}}, \ and\
  \bibinfo {author} {\bibfnamefont {H.~J.}\ \bibnamefont {Carmichael}},\ }\href
  {\doibase 10.1143/PTPS.64.307} {\bibfield  {journal} {\bibinfo  {journal}
  {Progress of Theoretical Physics Supplement}\ }\textbf {\bibinfo {volume}
  {64}},\ \bibinfo {pages} {307} (\bibinfo {year} {1978})}\BibitemShut
  {NoStop}%
\bibitem [{\citenamefont {Carmichael}(1980)}]{carm1980}%
  \BibitemOpen
  \bibfield  {author} {\bibinfo {author} {\bibfnamefont {H.~J.}\ \bibnamefont
  {Carmichael}},\ }\href@noop {} {\bibfield  {journal} {\bibinfo  {journal} {J.
  Phys. B: Atom. Molec. Phys.}\ }\textbf {\bibinfo {volume} {13}},\ \bibinfo
  {pages} {3551} (\bibinfo {year} {1980})}\BibitemShut {NoStop}%
\bibitem [{\citenamefont {Drummond}(1980)}]{drumm}%
  \BibitemOpen
  \bibfield  {author} {\bibinfo {author} {\bibfnamefont {P.~D.}\ \bibnamefont
  {Drummond}},\ }\href {\doibase 10.1103/PhysRevA.22.1179} {\bibfield
  {journal} {\bibinfo  {journal} {Phys. Rev. A}\ }\textbf {\bibinfo {volume}
  {22}},\ \bibinfo {pages} {1179} (\bibinfo {year} {1980})}\BibitemShut
  {NoStop}%
\bibitem [{\citenamefont {Drummond}\ and\ \citenamefont
  {Hassan}(1980)}]{drummHassan}%
  \BibitemOpen
  \bibfield  {author} {\bibinfo {author} {\bibfnamefont {P.~D.}\ \bibnamefont
  {Drummond}}\ and\ \bibinfo {author} {\bibfnamefont {S.~S.}\ \bibnamefont
  {Hassan}},\ }\href {\doibase 10.1103/PhysRevA.22.662} {\bibfield  {journal}
  {\bibinfo  {journal} {Phys. Rev. A}\ }\textbf {\bibinfo {volume} {22}},\
  \bibinfo {pages} {662} (\bibinfo {year} {1980})}\BibitemShut {NoStop}%
\bibitem [{\citenamefont {Drummond}\ and\ \citenamefont
  {Carmichael}(1978)}]{drummCarm}%
  \BibitemOpen
  \bibfield  {author} {\bibinfo {author} {\bibfnamefont {P.~D.}\ \bibnamefont
  {Drummond}}\ and\ \bibinfo {author} {\bibfnamefont {H.~J.}\ \bibnamefont
  {Carmichael}},\ }\href@noop {} {\bibfield  {journal} {\bibinfo  {journal}
  {Optics Communications}\ }\textbf {\bibinfo {volume} {27}} (\bibinfo {year}
  {1978})}\BibitemShut {NoStop}%
\bibitem [{\citenamefont {Barberena}\ \emph {et~al.}(2019)\citenamefont
  {Barberena}, \citenamefont {Lewis-Swan}, \citenamefont {Thompson},\ and\
  \citenamefont {Rey}}]{diego}%
  \BibitemOpen
  \bibfield  {author} {\bibinfo {author} {\bibfnamefont {D.}~\bibnamefont
  {Barberena}}, \bibinfo {author} {\bibfnamefont {R.~J.}\ \bibnamefont
  {Lewis-Swan}}, \bibinfo {author} {\bibfnamefont {J.~K.}\ \bibnamefont
  {Thompson}}, \ and\ \bibinfo {author} {\bibfnamefont {A.~M.}\ \bibnamefont
  {Rey}},\ }\href {\doibase 10.1103/PhysRevA.99.053411} {\bibfield  {journal}
  {\bibinfo  {journal} {Phys. Rev. A}\ }\textbf {\bibinfo {volume} {99}},\
  \bibinfo {pages} {053411} (\bibinfo {year} {2019})}\BibitemShut {NoStop}%
\bibitem [{\citenamefont {Morrison}\ and\ \citenamefont
  {Parkins}(2008)}]{Morrison_2008}%
  \BibitemOpen
  \bibfield  {author} {\bibinfo {author} {\bibfnamefont {S.}~\bibnamefont
  {Morrison}}\ and\ \bibinfo {author} {\bibfnamefont {A.~S.}\ \bibnamefont
  {Parkins}},\ }\href {\doibase 10.1088/0953-4075/41/19/195502} {\bibfield
  {journal} {\bibinfo  {journal} {Journal of Physics B: Atomic, Molecular and
  Optical Physics}\ }\textbf {\bibinfo {volume} {41}},\ \bibinfo {pages}
  {195502} (\bibinfo {year} {2008})}\BibitemShut {NoStop}%
\bibitem [{\citenamefont {Li}\ and\ \citenamefont {Paraoanu}(2009)}]{Li_2009}%
  \BibitemOpen
  \bibfield  {author} {\bibinfo {author} {\bibfnamefont {J.}~\bibnamefont
  {Li}}\ and\ \bibinfo {author} {\bibfnamefont {G.~S.}\ \bibnamefont
  {Paraoanu}},\ }\href {\doibase 10.1088/1367-2630/11/11/113020} {\bibfield
  {journal} {\bibinfo  {journal} {New Journal of Physics}\ }\textbf {\bibinfo
  {volume} {11}},\ \bibinfo {pages} {113020} (\bibinfo {year}
  {2009})}\BibitemShut {NoStop}%
\bibitem [{\citenamefont {Lee}\ \emph {et~al.}(2014{\natexlab{a}})\citenamefont
  {Lee}, \citenamefont {Reiter},\ and\ \citenamefont
  {Moiseyev}}]{Florentin_2014}%
  \BibitemOpen
  \bibfield  {author} {\bibinfo {author} {\bibfnamefont {T.~E.}\ \bibnamefont
  {Lee}}, \bibinfo {author} {\bibfnamefont {F.}~\bibnamefont {Reiter}}, \ and\
  \bibinfo {author} {\bibfnamefont {N.}~\bibnamefont {Moiseyev}},\ }\href
  {\doibase 10.1103/PhysRevLett.113.250401} {\bibfield  {journal} {\bibinfo
  {journal} {Phys. Rev. Lett.}\ }\textbf {\bibinfo {volume} {113}},\ \bibinfo
  {pages} {250401} (\bibinfo {year} {2014}{\natexlab{a}})}\BibitemShut
  {NoStop}%
\bibitem [{\citenamefont {Lee}\ and\ \citenamefont {Chan}(2013)}]{Lee_2013}%
  \BibitemOpen
  \bibfield  {author} {\bibinfo {author} {\bibfnamefont {T.~E.}\ \bibnamefont
  {Lee}}\ and\ \bibinfo {author} {\bibfnamefont {C.-K.}\ \bibnamefont {Chan}},\
  }\href {\doibase 10.1103/PhysRevA.88.063811} {\bibfield  {journal} {\bibinfo
  {journal} {Phys. Rev. A}\ }\textbf {\bibinfo {volume} {88}},\ \bibinfo
  {pages} {063811} (\bibinfo {year} {2013})}\BibitemShut {NoStop}%
\bibitem [{\citenamefont {Schneider}\ and\ \citenamefont
  {Milburn}(2002)}]{Milburn_2002}%
  \BibitemOpen
  \bibfield  {author} {\bibinfo {author} {\bibfnamefont {S.}~\bibnamefont
  {Schneider}}\ and\ \bibinfo {author} {\bibfnamefont {G.~J.}\ \bibnamefont
  {Milburn}},\ }\href {\doibase 10.1103/PhysRevA.65.042107} {\bibfield
  {journal} {\bibinfo  {journal} {Phys. Rev. A}\ }\textbf {\bibinfo {volume}
  {65}},\ \bibinfo {pages} {042107} (\bibinfo {year} {2002})}\BibitemShut
  {NoStop}%
\bibitem [{\citenamefont {Gonz\'alez-Tudela}\ and\ \citenamefont
  {Porras}(2013)}]{Porras_2013}%
  \BibitemOpen
  \bibfield  {author} {\bibinfo {author} {\bibfnamefont {A.}~\bibnamefont
  {Gonz\'alez-Tudela}}\ and\ \bibinfo {author} {\bibfnamefont {D.}~\bibnamefont
  {Porras}},\ }\href {\doibase 10.1103/PhysRevLett.110.080502} {\bibfield
  {journal} {\bibinfo  {journal} {Phys. Rev. Lett.}\ }\textbf {\bibinfo
  {volume} {110}},\ \bibinfo {pages} {080502} (\bibinfo {year}
  {2013})}\BibitemShut {NoStop}%
\bibitem [{\citenamefont {Lee}\ \emph {et~al.}(2014{\natexlab{b}})\citenamefont
  {Lee}, \citenamefont {Chan},\ and\ \citenamefont {Yelin}}]{Lee_2014}%
  \BibitemOpen
  \bibfield  {author} {\bibinfo {author} {\bibfnamefont {T.~E.}\ \bibnamefont
  {Lee}}, \bibinfo {author} {\bibfnamefont {C.-K.}\ \bibnamefont {Chan}}, \
  and\ \bibinfo {author} {\bibfnamefont {S.~F.}\ \bibnamefont {Yelin}},\ }\href
  {\doibase 10.1103/PhysRevA.90.052109} {\bibfield  {journal} {\bibinfo
  {journal} {Phys. Rev. A}\ }\textbf {\bibinfo {volume} {90}},\ \bibinfo
  {pages} {052109} (\bibinfo {year} {2014}{\natexlab{b}})}\BibitemShut
  {NoStop}%
\bibitem [{\citenamefont {Wolfe}\ and\ \citenamefont
  {Yelin}(2014)}]{Wolfe_2014}%
  \BibitemOpen
  \bibfield  {author} {\bibinfo {author} {\bibfnamefont {E.}~\bibnamefont
  {Wolfe}}\ and\ \bibinfo {author} {\bibfnamefont {S.~F.}\ \bibnamefont
  {Yelin}},\ }\href@noop {} {\enquote {\bibinfo {title} {Spin squeezing by
  means of driven superradiance},}\ } (\bibinfo {year} {2014}),\ \Eprint
  {http://arxiv.org/abs/arXiv:1405.5288} {arXiv:1405.5288} \BibitemShut
  {NoStop}%
\bibitem [{\citenamefont {Shankar}\ \emph {et~al.}(2017)\citenamefont
  {Shankar}, \citenamefont {Cooper}, \citenamefont {Bohnet}, \citenamefont
  {Bollinger},\ and\ \citenamefont {Holland}}]{Shankar_2017}%
  \BibitemOpen
  \bibfield  {author} {\bibinfo {author} {\bibfnamefont {A.}~\bibnamefont
  {Shankar}}, \bibinfo {author} {\bibfnamefont {J.}~\bibnamefont {Cooper}},
  \bibinfo {author} {\bibfnamefont {J.~G.}\ \bibnamefont {Bohnet}}, \bibinfo
  {author} {\bibfnamefont {J.~J.}\ \bibnamefont {Bollinger}}, \ and\ \bibinfo
  {author} {\bibfnamefont {M.}~\bibnamefont {Holland}},\ }\href {\doibase
  10.1103/PhysRevA.95.033423} {\bibfield  {journal} {\bibinfo  {journal} {Phys.
  Rev. A}\ }\textbf {\bibinfo {volume} {95}},\ \bibinfo {pages} {033423}
  (\bibinfo {year} {2017})}\BibitemShut {NoStop}%
\bibitem [{\citenamefont {Fink}\ \emph {et~al.}(2009)\citenamefont {Fink},
  \citenamefont {Bianchetti}, \citenamefont {Baur}, \citenamefont {G\"oppl},
  \citenamefont {Steffen}, \citenamefont {Filipp}, \citenamefont {Leek},
  \citenamefont {Blais},\ and\ \citenamefont {Wallraff}}]{Fink}%
  \BibitemOpen
  \bibfield  {author} {\bibinfo {author} {\bibfnamefont {J.~M.}\ \bibnamefont
  {Fink}}, \bibinfo {author} {\bibfnamefont {R.}~\bibnamefont {Bianchetti}},
  \bibinfo {author} {\bibfnamefont {M.}~\bibnamefont {Baur}}, \bibinfo {author}
  {\bibfnamefont {M.}~\bibnamefont {G\"oppl}}, \bibinfo {author} {\bibfnamefont
  {L.}~\bibnamefont {Steffen}}, \bibinfo {author} {\bibfnamefont
  {S.}~\bibnamefont {Filipp}}, \bibinfo {author} {\bibfnamefont {P.~J.}\
  \bibnamefont {Leek}}, \bibinfo {author} {\bibfnamefont {A.}~\bibnamefont
  {Blais}}, \ and\ \bibinfo {author} {\bibfnamefont {A.}~\bibnamefont
  {Wallraff}},\ }\href {\doibase 10.1103/PhysRevLett.103.083601} {\bibfield
  {journal} {\bibinfo  {journal} {Phys. Rev. Lett.}\ }\textbf {\bibinfo
  {volume} {103}},\ \bibinfo {pages} {083601} (\bibinfo {year}
  {2009})}\BibitemShut {NoStop}%
\bibitem [{\citenamefont {Mlynek}\ \emph {et~al.}(2014)\citenamefont {Mlynek},
  \citenamefont {Abdumalikov}, \citenamefont {Eichler},\ and\ \citenamefont
  {Wallraff}}]{Mlynek2014}%
  \BibitemOpen
  \bibfield  {author} {\bibinfo {author} {\bibfnamefont {J.~A.}\ \bibnamefont
  {Mlynek}}, \bibinfo {author} {\bibfnamefont {A.~A.}\ \bibnamefont
  {Abdumalikov}}, \bibinfo {author} {\bibfnamefont {C.}~\bibnamefont
  {Eichler}}, \ and\ \bibinfo {author} {\bibfnamefont {A.}~\bibnamefont
  {Wallraff}},\ }\href {\doibase 10.1038/ncomms6186} {\bibfield  {journal}
  {\bibinfo  {journal} {Nature Communications}\ }\textbf {\bibinfo {volume}
  {5}},\ \bibinfo {pages} {5186} (\bibinfo {year} {2014})}\BibitemShut
  {NoStop}%
\bibitem [{\citenamefont {Bonifacio}\ \emph {et~al.}(1971)\citenamefont
  {Bonifacio}, \citenamefont {Schwendimann},\ and\ \citenamefont
  {Haake}}]{Haake_1971}%
  \BibitemOpen
  \bibfield  {author} {\bibinfo {author} {\bibfnamefont {R.}~\bibnamefont
  {Bonifacio}}, \bibinfo {author} {\bibfnamefont {P.}~\bibnamefont
  {Schwendimann}}, \ and\ \bibinfo {author} {\bibfnamefont {F.}~\bibnamefont
  {Haake}},\ }\href {\doibase 10.1103/PhysRevA.4.302} {\bibfield  {journal}
  {\bibinfo  {journal} {Phys. Rev. A}\ }\textbf {\bibinfo {volume} {4}},\
  \bibinfo {pages} {302} (\bibinfo {year} {1971})}\BibitemShut {NoStop}%
\bibitem [{\citenamefont {Radcliffe}(1971)}]{Radcliffe_1971}%
  \BibitemOpen
  \bibfield  {author} {\bibinfo {author} {\bibfnamefont {J.~M.}\ \bibnamefont
  {Radcliffe}},\ }\href {\doibase 10.1088/0305-4470/4/3/009} {\bibfield
  {journal} {\bibinfo  {journal} {Journal of Physics A: General Physics}\
  }\textbf {\bibinfo {volume} {4}},\ \bibinfo {pages} {313} (\bibinfo {year}
  {1971})}\BibitemShut {NoStop}%
\bibitem [{\citenamefont {Wineland}\ \emph {et~al.}(1992)\citenamefont
  {Wineland}, \citenamefont {Bollinger}, \citenamefont {Itano}, \citenamefont
  {Moore},\ and\ \citenamefont {Heinzen}}]{Wineland_1992}%
  \BibitemOpen
  \bibfield  {author} {\bibinfo {author} {\bibfnamefont {D.~J.}\ \bibnamefont
  {Wineland}}, \bibinfo {author} {\bibfnamefont {J.~J.}\ \bibnamefont
  {Bollinger}}, \bibinfo {author} {\bibfnamefont {W.~M.}\ \bibnamefont
  {Itano}}, \bibinfo {author} {\bibfnamefont {F.~L.}\ \bibnamefont {Moore}}, \
  and\ \bibinfo {author} {\bibfnamefont {D.~J.}\ \bibnamefont {Heinzen}},\
  }\href {\doibase 10.1103/PhysRevA.46.R6797} {\bibfield  {journal} {\bibinfo
  {journal} {Phys. Rev. A}\ }\textbf {\bibinfo {volume} {46}},\ \bibinfo
  {pages} {R6797} (\bibinfo {year} {1992})}\BibitemShut {NoStop}%
\bibitem [{\citenamefont {Kitagawa}\ and\ \citenamefont {Ueda}(1993)}]{KU}%
  \BibitemOpen
  \bibfield  {author} {\bibinfo {author} {\bibfnamefont {M.}~\bibnamefont
  {Kitagawa}}\ and\ \bibinfo {author} {\bibfnamefont {M.}~\bibnamefont
  {Ueda}},\ }\href {\doibase 10.1103/PhysRevA.47.5138} {\bibfield  {journal}
  {\bibinfo  {journal} {Phys. Rev. A}\ }\textbf {\bibinfo {volume} {47}},\
  \bibinfo {pages} {5138} (\bibinfo {year} {1993})}\BibitemShut {NoStop}%
\bibitem [{\citenamefont {Zhang}\ \emph {et~al.}(2018)\citenamefont {Zhang},
  \citenamefont {Zhang},\ and\ \citenamefont {Molmer}}]{mcwf}%
  \BibitemOpen
  \bibfield  {author} {\bibinfo {author} {\bibfnamefont {Y.}~\bibnamefont
  {Zhang}}, \bibinfo {author} {\bibfnamefont {Y.}~\bibnamefont {Zhang}}, \ and\
  \bibinfo {author} {\bibfnamefont {K.}~\bibnamefont {Molmer}},\ }\href@noop {}
  {\bibfield  {journal} {\bibinfo  {journal} {New Journal of Physics}\ }\textbf
  {\bibinfo {volume} {20}} (\bibinfo {year} {2018})}\BibitemShut {NoStop}%
\bibitem [{\citenamefont {M{\o}lmer}\ \emph {et~al.}(1993)\citenamefont
  {M{\o}lmer}, \citenamefont {Castin},\ and\ \citenamefont {Dalibard}}]{mcwf2}%
  \BibitemOpen
  \bibfield  {author} {\bibinfo {author} {\bibfnamefont {K.}~\bibnamefont
  {M{\o}lmer}}, \bibinfo {author} {\bibfnamefont {Y.}~\bibnamefont {Castin}}, \
  and\ \bibinfo {author} {\bibfnamefont {J.}~\bibnamefont {Dalibard}},\ }\href
  {\doibase 10.1364/JOSAB.10.000524} {\bibfield  {journal} {\bibinfo  {journal}
  {J. Opt. Soc. Am. B}\ }\textbf {\bibinfo {volume} {10}},\ \bibinfo {pages}
  {524} (\bibinfo {year} {1993})}\BibitemShut {NoStop}%
\bibitem [{\citenamefont {Plenio}\ and\ \citenamefont {Knight}(1998)}]{mcwf3}%
  \BibitemOpen
  \bibfield  {author} {\bibinfo {author} {\bibfnamefont {M.~B.}\ \bibnamefont
  {Plenio}}\ and\ \bibinfo {author} {\bibfnamefont {P.~L.}\ \bibnamefont
  {Knight}},\ }\href {\doibase 10.1103/RevModPhys.70.101} {\bibfield  {journal}
  {\bibinfo  {journal} {Rev. Mod. Phys.}\ }\textbf {\bibinfo {volume} {70}},\
  \bibinfo {pages} {101} (\bibinfo {year} {1998})}\BibitemShut {NoStop}%
\bibitem [{SM()}]{SM}%
  \BibitemOpen
  \href@noop {} {\enquote {\bibinfo {title} {{See Supplemental Material at [URL
  will be inserted by publisher].}}}\ }\BibitemShut {NoStop}%
\bibitem [{\citenamefont {Shammah}\ \emph {et~al.}(2018)\citenamefont
  {Shammah}, \citenamefont {Ahmed}, \citenamefont {Lambert}, \citenamefont
  {De~Liberato},\ and\ \citenamefont {Nori}}]{Shammah_2018}%
  \BibitemOpen
  \bibfield  {author} {\bibinfo {author} {\bibfnamefont {N.}~\bibnamefont
  {Shammah}}, \bibinfo {author} {\bibfnamefont {S.}~\bibnamefont {Ahmed}},
  \bibinfo {author} {\bibfnamefont {N.}~\bibnamefont {Lambert}}, \bibinfo
  {author} {\bibfnamefont {S.}~\bibnamefont {De~Liberato}}, \ and\ \bibinfo
  {author} {\bibfnamefont {F.}~\bibnamefont {Nori}},\ }\href {\doibase
  10.1103/PhysRevA.98.063815} {\bibfield  {journal} {\bibinfo  {journal} {Phys.
  Rev. A}\ }\textbf {\bibinfo {volume} {98}},\ \bibinfo {pages} {063815}
  (\bibinfo {year} {2018})}\BibitemShut {NoStop}%
\bibitem [{\citenamefont {Norcia}\ \emph {et~al.}(2016)\citenamefont {Norcia},
  \citenamefont {Winchester}, \citenamefont {Cline},\ and\ \citenamefont
  {Thompson}}]{Norcia_2016}%
  \BibitemOpen
  \bibfield  {author} {\bibinfo {author} {\bibfnamefont {M.~A.}\ \bibnamefont
  {Norcia}}, \bibinfo {author} {\bibfnamefont {M.~N.}\ \bibnamefont
  {Winchester}}, \bibinfo {author} {\bibfnamefont {J.~R.~K.}\ \bibnamefont
  {Cline}}, \ and\ \bibinfo {author} {\bibfnamefont {J.~K.}\ \bibnamefont
  {Thompson}},\ }\href {\doibase 10.1126/sciadv.1601231} {\bibfield  {journal}
  {\bibinfo  {journal} {Science Advances}\ }\textbf {\bibinfo {volume} {2}}
  (\bibinfo {year} {2016}),\ 10.1126/sciadv.1601231}\BibitemShut {NoStop}%
\bibitem [{\citenamefont {Hutson}\ \emph {et~al.}(2019)\citenamefont {Hutson},
  \citenamefont {Goban}, \citenamefont {Marti}, \citenamefont {Sonderhouse},
  \citenamefont {Sanner},\ and\ \citenamefont {Ye}}]{Hutson2019}%
  \BibitemOpen
  \bibfield  {author} {\bibinfo {author} {\bibfnamefont {R.~B.}\ \bibnamefont
  {Hutson}}, \bibinfo {author} {\bibfnamefont {A.}~\bibnamefont {Goban}},
  \bibinfo {author} {\bibfnamefont {G.~E.}\ \bibnamefont {Marti}}, \bibinfo
  {author} {\bibfnamefont {L.}~\bibnamefont {Sonderhouse}}, \bibinfo {author}
  {\bibfnamefont {C.}~\bibnamefont {Sanner}}, \ and\ \bibinfo {author}
  {\bibfnamefont {J.}~\bibnamefont {Ye}},\ }\href {\doibase
  10.1103/PhysRevLett.123.123401} {\bibfield  {journal} {\bibinfo  {journal}
  {Phys. Rev. Lett.}\ }\textbf {\bibinfo {volume} {123}},\ \bibinfo {pages}
  {123401} (\bibinfo {year} {2019})}\BibitemShut {NoStop}%
\end{thebibliography}%

\newpage 

\onecolumngrid
\vspace{\columnsep}
\begin{center}
\textbf{\large Supplemental Material: Single-Particle Decoherence Can Improve Spin-Squeezing Generated In Collective Dynamics}
\end{center}
\vspace{\columnsep}
\twocolumngrid

\setcounter{equation}{0}
\setcounter{figure}{0}
\setcounter{table}{0}
\setcounter{page}{1}
\makeatletter
\renewcommand{\theequation}{S\arabic{equation}}
\renewcommand{\thefigure}{S\arabic{figure}}

In this supplemental material we summarize supporting details related to the numerical analysis used in the manuscript. We cover the mean-field analysis in Sec.~\ref{sec:MF}, a cumulant expansion in Sec.~\ref{sec:Cumulant} and briefly summarize a Monte Carlo wavefunction approach in Sec.~\ref{sec:MC}. 

\section{Mean-Field Analysis \label{sec:MF}}
We use a mean-field approach to analyse the system for $\gamma_s > 0$, for which no analytic expression steady-state exists. This enables us to 
gain insight into the steady-state phase diagram of the system. Moreover, we use numerical simulations of the master equation for small system sizes ($N\sim 1000$) to confirm the feasibility of the mean field predictions. For completeness, the master equation investigated in the main text is given by
\begin{flalign}
\label{eqn_master}
\frac{\partial\hat{\rho}}{\partial t} &= -\frac{i}{\hbar}[\hat{H},\hat{\rho}] + L_c[\hat{\rho}] + L_s[\hat{\rho}], \\
\hat{H} &= \chi \hat{J}_+\hat{J}_- + \Omega \hat{J}_x . \nonumber
\end{flalign}
In the remainder of this supplemental material we set $\hbar = 1$. 
 
In the mean-field approximation, we derive equations of motion for the expectations of individual particle Pauli operators, $\partial_t\est{i}{\alpha} = \mbox{Tr}[\sigma_{i}^{\alpha}\partial_t\hat{\rho}]$, from the master equation [Eq.~(\ref{eqn_master})]. Due to the permutational symmetry of the master equation, and assuming the same symmetry applies to the initial state, these  equations will be identical for all particles.

Under the mean-field assumption $\rho = \bigotimes_i \rho_i$, where each $\rho_i$ is a single particle density, the equations can be closed as second-order expectations can be factored as $\esst{i}{\alpha}{j}{\beta} = \est{i}{\alpha}\est{j}{\beta}$ when $i \neq j$. When $i = j$, second order expectations can be handled in one of two ways. The first approach would be to use commutation relations to resolve the product of Pauli operators into a single operator before taking the expected value, i.e. $\langle \hat{\sigma}^{\alpha}_i \hat{\sigma}^{\beta}_i \rangle \to \delta_{\alpha,\beta} +  i\epsilon_{\alpha\beta\gamma}\langle \hat{\sigma}^{\gamma}_i \rangle$. The second approach is to factor in the same way as for unlike particles, i.e. $\langle \hat{\sigma}^{\alpha}_i \hat{\sigma}^{\beta}_i \rangle \to \langle \hat{\sigma}^{\alpha}_i \rangle \langle \hat{\sigma}^{\beta}_i \rangle$. Previous work \cite{carm1980} has shown that the former approach aligns with the exact solution when $\hat{J}^2$ is not conserved, and that the latter is accurate otherwise. Since we are interested in the case where $\gamma_s > 0$ and total spin is not conserved, we factor only unlike particles. Dropping the subscripts due to particle symmetry and defining $\ev{\sigma^+} \equiv r e^{i\phi}$, and $\ev{\sigma_z} \equiv z$, we arrive at the mean-field equations
\begin{flalign}
\label{eqn_mf1}
\dot{r} = & -\frac{\Gamma + \gamma_s}{2}r + \frac{\Gamma}{2}(N - 1) z\, r -\frac{\Omega}{2} z \sin{\phi} \\
\dot{\phi} = & -\chi(N - 1)z - \frac{\Omega}{2r} z \cos{\phi} + \chi \\
\dot{z} = & -2\Gamma(N - 1) r^2 - (\Gamma + \gamma_s)(1 + z)  + 2\Omega \, r \sin{\phi}. 
\end{flalign}

We determine the steady-state at the mean-field level by setting the LHS of Eqs.~(\ref{eqn_mf1}) to zero and solving for $(r,\phi,z)$.
We begin by deriving a steady-state expression for $z$ in terms of the other variables. From the $r$ equation, we get
\[ 0 = -\frac{\Gamma + \gamma_s}{2}r + \frac{\Gamma(N - 1)}{2}z\,r - \frac{\Omega}{2} z \sin\phi\]
which implies that
\[ -2(N-1)\Gamma r^2 + 2\Omega\, r \sin\phi = -2(\Gamma + \gamma_s)\frac{r^2}{z}, \]
where we have assumed $z \neq 0$. Plugging this into the $z$ equation gives
\begin{flalign*}
 0 = & -2(\Gamma + \gamma_s)\frac{r^2}{z} - (\Gamma + \gamma_s)(1+z) \\
\Rightarrow 0 = & z^2 + z + 2r^2, 
\end{flalign*}
and subsequently,
\begin{equation}
\label{eqn_z}
z = -\frac{1}{2} \pm \frac{1}{2}\sqrt{1 - 8r^2}.
\end{equation}
Here, we identify that there are two steady-state values of $z$. A linear stability analysis reveals that the branch with larger absolute value is stable, while the smaller is unstable. We also see that we can expect a bifurcation when $r$ passes through $1/\sqrt{8}$ from below. 

We now turn our attention to the steady-state value of $r$. From the $z$ equation, we see that
\begin{equation}
\label{eqn_sinphi}
4r^2\Omega^2\sin^2\phi = [2(N-1)\Gamma\,r^2 + (\Gamma + \gamma_s)(1+z)]^2,
\end{equation}
and from the $\phi$ equation
\begin{flalign}
\label{eqn_cosphi}
4r^2\Omega^2\cos^2\phi = & 4r^2(1 - \sin^2\phi) \nonumber \\
= & \left[-\chi(N-1)r^2 + \chi \frac{r^2}{z}\right]^2.
\end{flalign}

\noindent
Combining Eqs.~(\ref{eqn_sinphi}) and (\ref{eqn_cosphi}) we get
\begin{flalign}
\label{eqn_r}
4r^2\Omega^2 - \left[ 2(N-1)\Gamma\,r^2 + (\Gamma + \gamma_s)(1+z) \right]^2  \nonumber \\
=  \left[-\chi(N-1)r^2 + \chi \frac{r^2}{z}\right]^2,
\end{flalign}
where $z$ is given in terms of $r$ by Eq.~(\ref{eqn_z}). Note that if we now restrict our attention to leading order in $N$, the above implies
\[ 4r^2\Omega^2 - 4N^2\Gamma^2r^4 = N^2\chi^2r^4\]
\begin{equation}
\label{eqn_r_largen}
\Rightarrow r^2 = \frac{\Omega^2}{N^2(\Gamma^2 + 4\chi^2)} = \frac{\Upsilon^2}{4\Upsilon_c^2},
\end{equation}
a familiar result that shows that we expect an increasing value of $r$ as we increase the Rabi frequency of the drive $\Omega$. Recall, however, that Eq.~(\ref{eqn_z}) predicts a phase transition when $r = 1/\sqrt{8}$. This corresponds to 
\begin{equation}
\label{eqn_ocp}
\Upsilon = \frac{\Upsilon_c}{\sqrt{2}} \equiv \Upsilon_c'.
\end{equation}
Beyond this transition, equation (\ref{eqn_z}) predicts complex $z$, which is not a valid steady-state. Recall, however, that equation (\ref{eqn_z}) assumed $z \neq 0$. Numerical simulations confirm that for $\Upsilon > \Upsilon_c'$ the steady-state value of $z$ is, in fact, zero. This aligns with the transition between superradiant and normal phases that we were expecting. We will see that this transition exists beyond the mean-field level when we look at numerical solutions to the cumulant expansion equations in the next section.

\section{Cumulant Expansion \label{sec:Cumulant}}

\begin{figure*}[t!] 
   \centering
   \includegraphics[width=7.0cm]{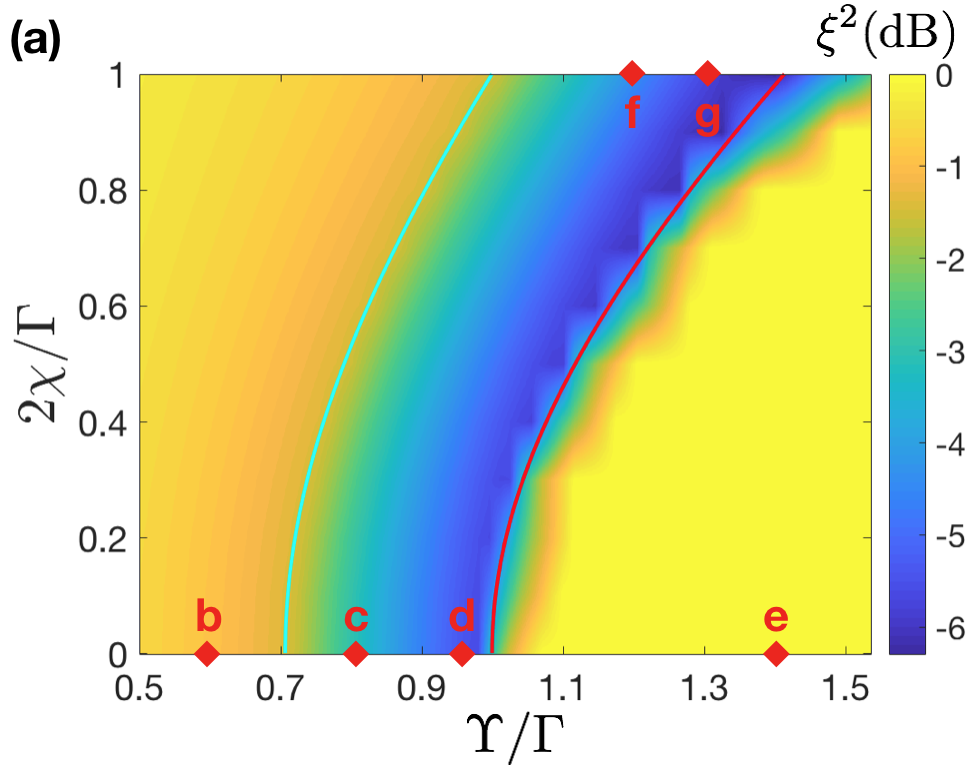}
   \includegraphics[width=6.9cm]{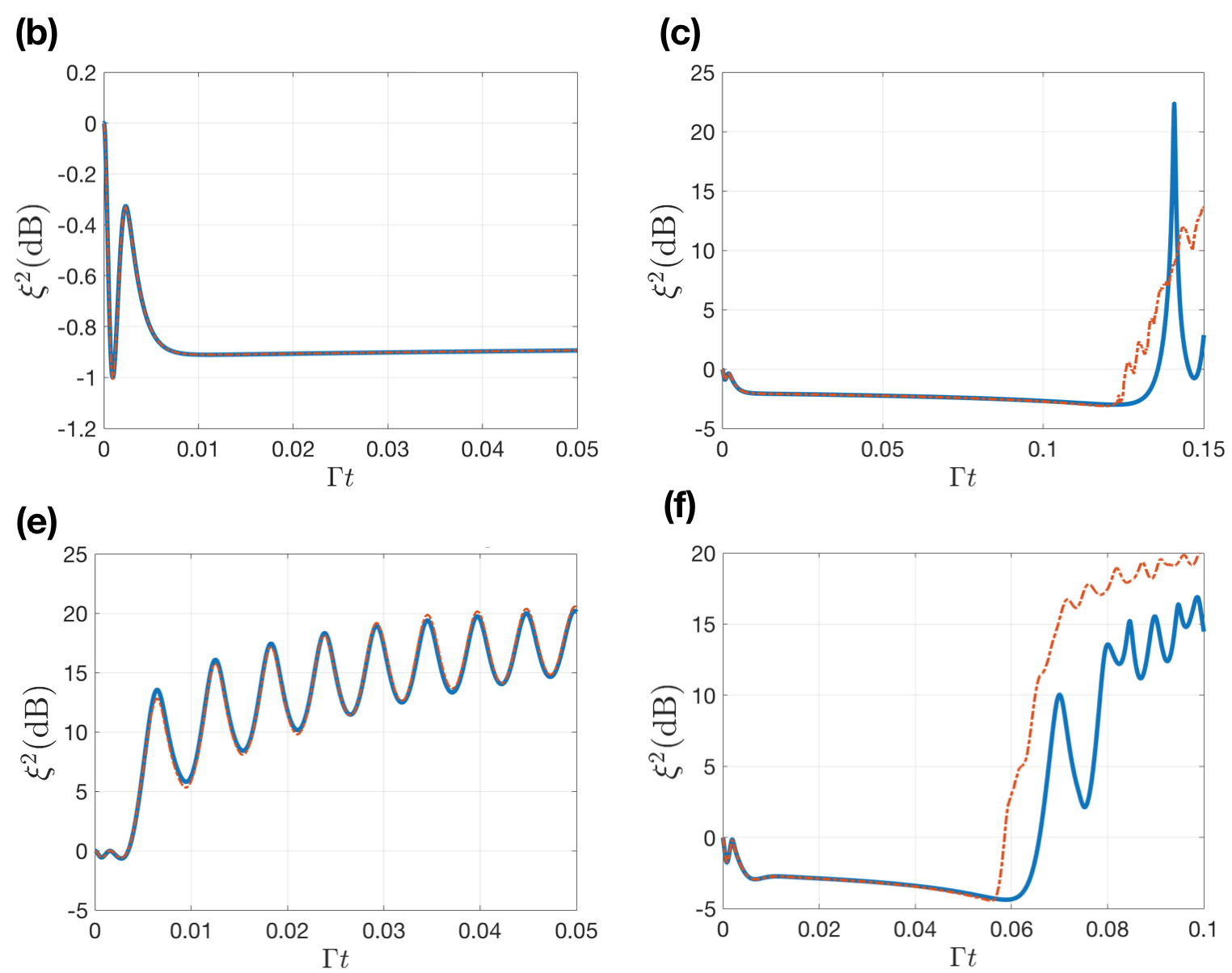} 
   \includegraphics[width=3.5cm]{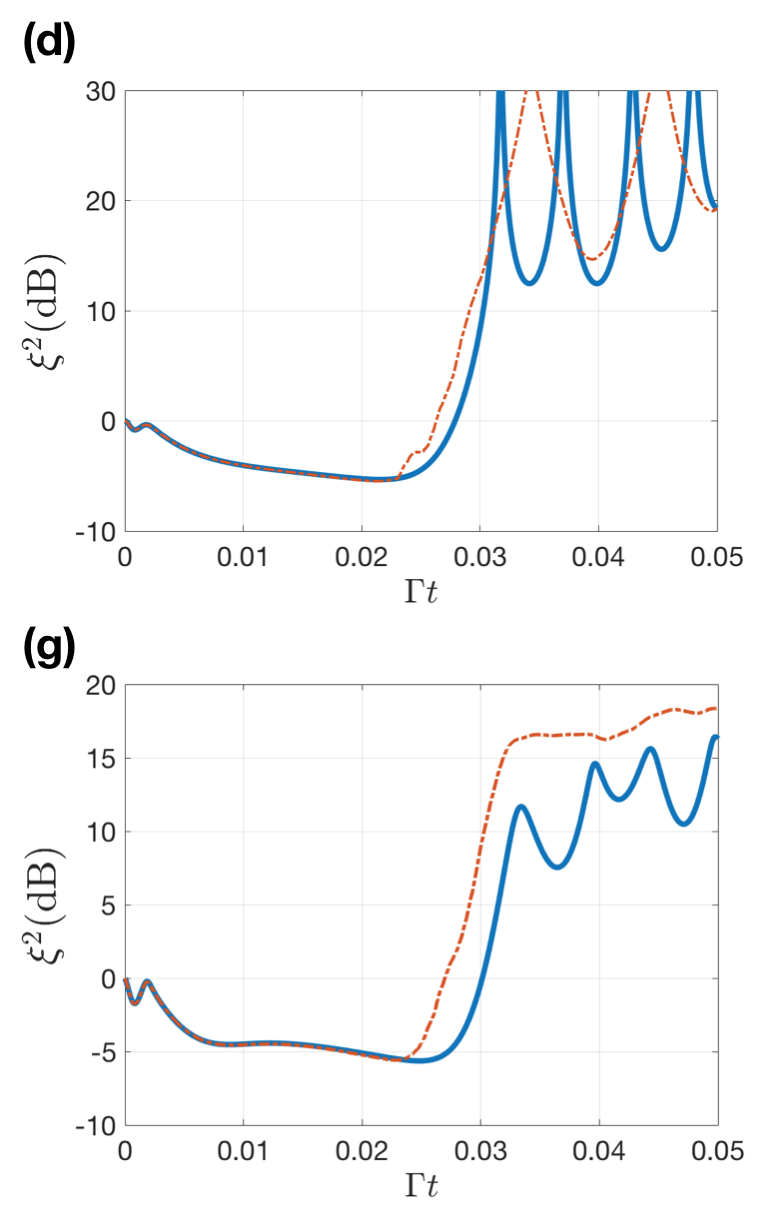} 
   \caption{(a) Minimum squeezing attained from cumulant model after $t_{\mbox{min}} = 0.003$ where $N = 2000$ and $\gamma_s/\Gamma = 20$ with the phase boundaries $\Upsilon_c'/\Gamma$ (blue) and $\Upsilon_c/\Gamma$ (red). (b)-(g) Comparison of squeezing dynamics between cumulant (blue) and MCWF (red) where $\chi/\Gamma$ and $2\Omega/N\Gamma$ values are shown in red in panel (a)}
   \label{sm_fig1}
\end{figure*}

To obtain a model that can be numerically integrated efficiently but also allows for quantum correlations, we turn our attention to a cumulant expansion approximation. Specifically, we extend the mean-field analysis by computing equations of motion for expectation values of products of two Pauli operators, e.g., $\partial_t\langle\hat{\sigma}^{\alpha}_i\hat{\sigma}^{\beta}_j\rangle = \mathrm{Tr}[\hat{\sigma}^{\alpha}_i\hat{\sigma}^{\beta}_j\partial_t\hat{\rho}]$, in addition to the equations of motion for the expectations of individual Pauli operators. Together these form a hierarchy of equations which we truncate by assuming that third order expectations can be factorized into products of lower-order terms. Specifically, we assume they factorize according to
\begin{flalign}
\label{eqn_cumulant_exp}
\langle\hat{\sigma}_a^\alpha\hat{\sigma}_b^\beta\hat{\sigma}_c^\gamma\rangle \approx & \,\langle\hat{\sigma}_a^\alpha\hat{\sigma}_b^\beta\rangle\langle\hat{\sigma}_c^\gamma\rangle+\langle\hat{\sigma}_b^\beta\hat{\sigma}_c^\gamma\rangle\langle\hat{\sigma}_a^\alpha\rangle+\langle\hat{\sigma}_a^\alpha\hat{\sigma}_c^\gamma\rangle\langle\hat{\sigma}_b^\beta\rangle \nonumber \\
- & 2\langle\hat{\sigma}_a^\alpha\rangle\langle\hat{\sigma}_b^\beta\rangle\langle\hat{\sigma}_c^\gamma\rangle,
\end{flalign}
in the case where $a, b$ and $c$ are distinct. Operator products for the same particle are resolved using the usual Pauli relations. Combining this factorization with the particle symmetry present in the master equation, and assuming all particles start with the same initial conditions, this results in a closed system of six complex ordinary differential equations:
\begin{widetext}
\begin{eqnarray}
\frac{d\langle\hat{\sigma}_a^+\rangle}{dt}=&&-\frac{\Gamma +\gamma_s}{2}\langle\hat{\sigma}_a^+\rangle+\frac{1}{2}(N-1)(\Gamma-i2\chi)\langle\hat{\sigma}_a^z\hat{\sigma}_b^+\rangle -i\frac{\Omega}{2}\es{a}{z} + i\chi\es{a}{+},\\
\frac{d\langle\hat{\sigma}_a^z\hat{\sigma}_b^+\rangle}{dt}=&&-\frac{3}{2}(\Gamma +\gamma_s)\langle\hat{\sigma}_a^z\hat{\sigma}_b^+\rangle-(\Gamma +\gamma_s)\langle\hat{\sigma}_b^+\rangle-\frac{1}{2}(\Gamma+i2\chi)\langle\hat{\sigma}_a^+\rangle-\Gamma \langle\hat{\sigma}_a^+\hat{\sigma}_b^z\rangle\nonumber\\
&&+\frac{1}{2}(N-2)(\Gamma-i2\chi)\langle\hat{\sigma}_a^z\hat{\sigma}_b^z\hat{\sigma}_j^+\rangle-(N-2)(\Gamma+i2\chi)\langle\hat{\sigma}_a^+\hat{\sigma}_b^+\hat{\sigma}_j^-\rangle-(N-2)(\Gamma-i2\chi)\langle\hat{\sigma}_a^-\hat{\sigma}_b^+\hat{\sigma}_j^+\rangle \nonumber \\
&& -i\frac{\Omega}{2}\left(2(\ess{a}{+}{b}{+} - \ess{a}{-}{b}{+}) + \ess{a}{z}{b}{z}\right) + i\chi\ess{a}{z}{b}{+},\\
\frac{d\langle\hat{\sigma}_a^z\rangle}{dt}=&&-2i\chi(N-1)(\langle \hat{\sigma}_a^+\hat{\sigma}_b^-\rangle-\langle \hat{\sigma}_a^-\hat{\sigma}_b^+\rangle)-\Gamma(N-1)(\langle\hat{\sigma}_a^+\hat{\sigma}_b^-\rangle+\langle\hat{\sigma}_a^-\hat{\sigma}_b^+\rangle)-(\Gamma + \gamma_s)(\langle\hat{\sigma}_a^z\rangle + 1) \nonumber \\
&&+ 2\Omega\mbox{Im}[\es{a}{+}],\\
\frac{d\langle \hat{\sigma}_a^+\hat{\sigma}_b^-\rangle}{dt}=&&\frac{1}{2}(N-2)(\Gamma-i2\chi)\langle\hat{\sigma}_a^z\hat{\sigma}_b^-\hat{\sigma}_j^+\rangle+\frac{1}{2}(N-2)(\Gamma+i2\chi)\langle\hat{\sigma}_b^z\hat{\sigma}_a^+\hat{\sigma}_j^-\rangle +\frac{\Gamma}{2}\left(\langle\hat{\sigma}_a^z\hat{\sigma}_b^z\rangle+\langle\hat{\sigma}_a^z\rangle\right) \nonumber\\
&&-(\Gamma + \gamma_s) \langle\hat{\sigma}_a^+\hat{\sigma}_b^-\rangle - \Omega\mbox{Im}[\ess{a}{z}{b}{+}],\\
\frac{d\langle\hat{\sigma}_a^z\hat{\sigma}_b^z\rangle}{dt}=&&-i(N-2)2\chi[\langle\hat{\sigma}_a^+\hat{\sigma}_b^z\hat{\sigma}_j^-\rangle-\langle\hat{\sigma}_a^-\hat{\sigma}_b^z\hat{\sigma}_j^+\rangle]-i(N-2)2\chi[\langle\hat{\sigma}_b^+\hat{\sigma}_a^z\hat{\sigma}_j^-\rangle-\langle\hat{\sigma}_b^-\hat{\sigma}_a^z\hat{\sigma}_j^+\rangle]\nonumber\\
&&-(N-2)\Gamma[\langle\hat{\sigma}_a^+\hat{\sigma}_b^z\hat{\sigma}_j^-\rangle
+\langle\hat{\sigma}_a^-\hat{\sigma}_b^z\hat{\sigma}_j^+\rangle]-(N-2)\Gamma[\langle\hat{\sigma}_b^+\hat{\sigma}_a^z\hat{\sigma}_j^-\rangle+\langle\hat{\sigma}_b^-\hat{\sigma}_a^z\hat{\sigma}_j^+\rangle]\nonumber\\
&&-2(\Gamma + \gamma_s)\langle\hat{\sigma}_a^z\rangle-2(\Gamma +\gamma_s)\langle\hat{\sigma}_a^z\hat{\sigma}_b^z\rangle+4\Gamma\mbox{Re}[ [\langle\hat{\sigma}_a^+\hat{\sigma}_b^-\rangle] + 4\Omega\mbox{Im}[\ess{a}{z}{b}{+}],\\
\frac{d\langle\hat{\sigma}_a^+\hat{\sigma}_b^+\rangle}{dt}=&&(N-2)(\Gamma-i2\chi)\langle\hat{\sigma}_a^z\hat{\sigma}_b^+\hat{\sigma}_j^+\rangle-(\Gamma + \gamma_s)\langle\hat{\sigma}_a^+\hat{\sigma}_b^+\rangle - i\Omega\ess{a}{z}{b}{+} + 2i\chi\ess{a}{+}{b}{+}.
\end{eqnarray}
\end{widetext}

We numerically integrate these cumulant equations to obtain the dynamics of the driven-dissipative system with $\gamma_s > 0$. In Fig.~\ref{sm_fig1}(a) we plot the best squeezing attained after a given threshold timescale (related to the early transient collective squeezing), for a range of $\chi/\Gamma$ and $2\Omega/N\Gamma$ values with $N = 2000$ and $\gamma_s/\Gamma = 20$
Note that minimum values greater than zero have been cut off at zero so as to avoid saturating the color plot. The two phase boundaries $\Omega_c$ and $\Omega_c'$ are clearly visible, bracketing the region where spontaneous emission enhanced squeezing is allowed to develop.

Panels (b)-(e) of Fig.~\ref{sm_fig1} show a comparison of results from the cumulant expansion along with those from the MCWF method (which is numerically exact in the limit of infinite trajectories, and further discussed in the following section). It can be seen that there is close agreement up to the crossing of the dynamical phase transition, making the result obtained from the cumulant equations a reasonable indicator of the true extent and timing of maximum squeezing.

\section{Exact Solver \label{sec:MC}}

\noindent
While the cumulant approximation is attractive as it allows us to obtain numerical results rapidly for a large system, there are also other numerical methods to solve the full quantum dynamics of the system in a reasonably efficient manner. In particular, Ref.~\cite{mcwf} recently demonstrated that a Monte Carlo wavefunction approach can be implemented to efficiently solve the dissipative dynamics of spin systems exhibiting permutational symmetry for $N \sim \mathcal{O}(10^3)$ and even up to $N \sim \mathcal{O}(10^5)$ in special cases. While we point the interested reader to Ref.~\cite{mcwf} for full details of the numerical method, we summarize here the key aspects and advantages. 


The MCWF method unravels the density matrix into an ensemble of pure state wave functions that evolve independently of one another in time, where dissipation is handled by random jumps. The full time evolution of one member of this ensemble is referred to as a trajectory. The time evolution of the density matrix is recovered by taking the average of the pure state density matrices at each point in time, resulting in the mixed state solution to the master equation [Eq.~(\ref{eqn_master})].

The advantages of this method are three-fold. First, as shown in \cite{mcwf}, even though spontaneous emission breaks the $\hat{J}^2 = \hat{J}_x^2 + \hat{J}_y^2 + \hat{J}_z^2$ symmetry in the master equation, each quantum trajectory lies within a single eigenspace of total spin at any given time. This means that each trajectory can be efficiently integrated, as the dimensionality of the Hilbert space in which it exists is only $\orderof(N)$. Second, the trajectories evolve in time independently of one another, allowing for the parallel simulation of different trajectories and thus rapid evaluation of ensemble averages. Finally, analyzing the time evolution of individual trajectories can provide insight that is not altogether obvious from the evolution of the density matrix resulting from the ensemble averages, as was discussed in the main text.

\begin{figure}[hbt] 
   \centering
   \includegraphics[width=7.0cm]{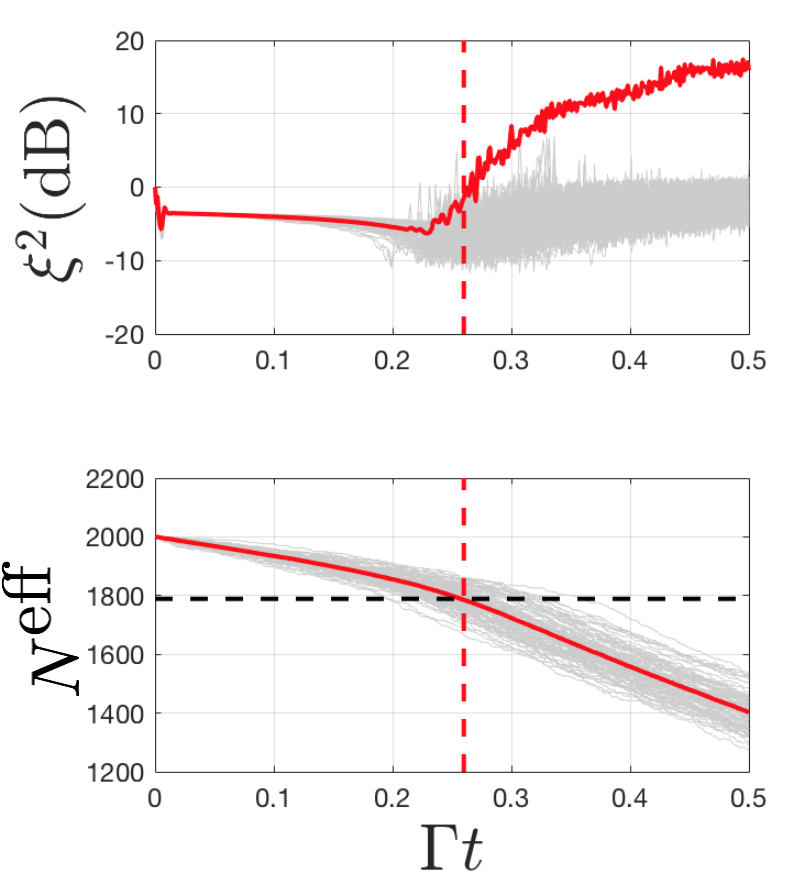}
   \caption{Spin-squeezing and effective particle number versus time for the full state (red) and a number of individual trajectories (grey) where $N = 2000$, $\Omega/\Gamma = 2000$, $\chi/\Gamma = 1$, and $\gamma_s/\Gamma = 2$. Initial conditions are the coherent spin state in the $-x$-direction. The black horizontal line indicates the critical $N^{\mbox{eff}}$ where $\Upsilon_c^{\mbox{eff}} = \Upsilon$ and the red vertical line is the point in time where it is crossed}
   \label{sm_fig2}
\end{figure}

Following the discussion in the main text, in Fig.~\ref{sm_fig2} we plot an example of the squeezing and effective particle number versus time from the ensemble average and also individual trajectories. We observe that the squeezing improves until enough trajectories approach the critical $N^{\mbox{eff}}$ threshold. After a sufficient number of trajectories cross the threshold and lose their squeezing, the overall squeezed state is then quickly lost. This critical number of trajectories can be reached even before the ensemble average $N^{\mbox{eff}}$ crosses the critical threshold, as is the case in the figure.

\end{document}